\RequirePackage{amsthm}
\documentclass[pdflatex,sn-mathphys-ay]{sn-jnl}

\usepackage{graphicx}%
\usepackage{multirow}%
\usepackage{amsmath,amssymb,amsfonts}%
\usepackage{amsthm}%
\usepackage{mathrsfs}%
\usepackage[title]{appendix}%
\usepackage{xcolor}%
\usepackage{textcomp}%
\usepackage{manyfoot}%
\usepackage{booktabs}%
\usepackage{algorithm}%
\usepackage{algorithmicx}%
\usepackage{algpseudocode}%
\usepackage{listings}%
\usepackage{pdflscape}

\usepackage{bm}
\usepackage{anyfontsize}

\global\long\def\HS{\text{\tiny HS}}
\global\long\def\BT{\text{\tiny BT}}
\global\long\def\F{\text{\tiny F}}
\global\long\def\N{\text{\tiny N}}
\global\long\def\B{\text{\tiny B}}
\global\long\def\QCZ{\text{\tiny QCZ}}
\global\long\def\ZZC{\text{\tiny ZZC}}
\global\long\def\NEW{\text{\tiny NEW}}
\global\long\def\SP{\operatorname{SP}}
\global\long\def\GP{\operatorname{GP}}
\global\long\def\tr{\operatorname{tr}}
\global\long\def\Cov{\operatorname{Cov}}
\global\long\def\ARE{\operatorname{ARE}}
\global\long\def\Var{\operatorname{Var}}
\global\long\def\diag{\operatorname{diag}}
\global\long\def\E{\operatorname{E}}
\global\long\def\iidsim{\stackrel{\text{i.i.d.}}{\sim}}
\global\long\def\dequ{\stackrel{d}{=}}
\global\long\def\calL{\mathcal{L}}
\global\long\def\convL{\stackrel{\calL}{\longrightarrow}}
\global\long\def\convd{\stackrel{d}{\longrightarrow}}
\global\long\def\convP{\stackrel{P}{\longrightarrow}}

\global\long\def\b#1{{\bf \bm{\mathit{#1}}}}
\global\long\def\bzero{\b 0}
\global\long\def\bfeta{\b{\eta}}

\global\long\def\bn{\b n}
\global\long\def\bx{\b x}
\global\long\def\by{\b y}
\global\long\def\bz{\b z}
\global\long\def\bGamma{\b{\Gamma}}
\global\long\def\bOmega{\b{\Omega}}
\global\long\def\barby{\bar{\by}}
\global\long\def\barbx{\bar{\bx}}
\global\long\def\bA{\b A}
\global\long\def\bB{\b B}
\global\long\def\bC{\b C}
\global\long\def\bD{\b D}
\global\long\def\bG{\b G}
\global\long\def\bbH{\b H}
\global\long\def\bI{\b I}
\global\long\def\bM{\b M}
\global\long\def\bP{\b P}

\global\long\def\bX{\b X}
\global\long\def\bphi{\b{\phi}}
\global\long\def\bDelta{\b{\Delta}}

\global\long\def\calT{\mathcal{T}}
\global\long\def\calK{\mathcal{K}}
\newcommand{\LpT}{{\cal L}^p_2(\cal T)}

\global\long\def\h#1{\widehat{#1}}
\global\long\def\hd{\h d}

\global\long\def\bSigma{\b{\Sigma}}
\global\long\def\hbeta{\h{\beta}}
\global\long\def\hbGamma{\h{\bGamma}}
\global\long\def\hbOmega{\h{\bOmega}}
\global\long\def\hbM{\h {\bM}}
\global\long\def\t#1{\widetilde{#1}}
\global\long\def\tbG{\t{\bG}}

\usepackage{soul,xcolor,cancel}
\usepackage{color}

\setstcolor{red}
\newif\ifApproveEdit
\ApproveEditfalse
\ifApproveEdit
\newcommand{\mrk}[1]{{#1}}
\newcommand{\del}[1]{\iffalse{#1}\fi}
\else
\newcommand{\mrk}[1]{\textcolor{red}{#1}}
\newcommand\del[2][red]{\setbox0=\hbox{$#2$}%
\rlap{\raisebox{.45\ht0}{\textcolor{#1}{\rule{\wd0}{1pt}}}}#2}
\fi

\theoremstyle{thmstyleone}%
\newtheorem{theorem}{Theorem}
\theoremstyle{thmstyletwo}%
\newtheorem{remark}{Remark}%

\theoremstyle{thmstylethree}%

\begin{document}

\title[The general linear hypothesis testing problem for multivariate functional data with applications]{The general linear hypothesis testing problem for multivariate functional data with applications}


\author*[1]{\fnm{Tianming} \sur{Zhu}}\email{tianming.zhu@nie.edu.sg}

\affil*[1]{\orgdiv{National Institute of Education}, \orgname{Nanyang Technological University}, \orgaddress{\street{1 Nanyang Walk}, \city{Singapore}, \postcode{637616},  \country{Singapore}}}


\abstract{As technology continues to advance at a rapid pace, the prevalence of multivariate functional data (MFD)  has expanded across diverse disciplines, spanning biology, climatology, finance, and numerous other fields of study. Although MFD are encountered in various fields, the development of methods for hypotheses on mean functions, especially the general linear hypothesis testing (GLHT) problem for such data has been limited. 
In this study, we propose and study a new global test for the GLHT problem for MFD, which includes the  one-way multivariate analysis of variance for functional data (FMANOVA), post hoc, and contrast analysis as special cases. The asymptotic null distribution of the test statistic is shown to be a chi-squared-type mixture dependent of eigenvalues of the heteroscedastic covariance functions. The distribution of the  chi-squared-type mixture can be well approximated by a three-cumulant matched chi-squared-approximation with its approximation parameters estimated from the data. By incorporating an adjustment coefficient, the proposed test performs effectively irrespective of the correlation structure in the functional data, even when dealing with a relatively small sample size. Additionally, the asymptotic power of the proposed test under a local alternative is established.     Simulation studies and a real data example demonstrate finite-sample performance and broad applicability of the proposed test. 
}




\maketitle

\section{Introduction}
\label{sec:intro}

With the rapid evolution of technology, data are increasingly being acquired and represented as trajectories or images in various scientific domains. This trend can be observed in disciplines such as meteorology, biology, medicine, and engineering, among others. Consequently, functional data analysis (FDA) has emerged as a prominent field of study with significant applications across numerous real-world domains.  It is a branch of statistics concerned with the analysis of infinite-dimensional variables and can be treated as a generalization of the standard multivariate data analysis. The work of this paper is partially motivated by a financial data set provided by the Credit Research Initiative of National University of Singapore (NUS-CRI). 
The Probability of Default (PD) quantifies the likelihood of an obligor defaulting on its financial obligations. It is a core credit product of the NUS-CRI corporate default prediction system, which is based on the forward intensity model of \cite{duan2012multiperiod}. This financial data set contains the  mean PD values aggregated by the economy of domicile and sector of each firm from 2012 to 2021. As a result,  each economy can be represented by multiple curves, with each curve representing the aggregate PD of a specific sector, illustrating  an example of multivariate functional observations. It is of interest to compare the mean aggregated PD curves corresponding to four important factors, namely, energy, financial, real estate, and industrial in the various regions, including Asia Pacific (Developed),  Asia Pacific (Emerging), Eurozone, and Non-Eurozone  are all the same. This leads to a $k$-sample problem for multivariate functional data (MFD)  which is a special case of the general linear hypothesis testing (GLHT) problem for MFD.

Mathematically, a general $k$-sample problem, also known as one-way multivariate analysis of variance for functional data (FMANOVA) can be described as follows. Let $\SP_p(\bfeta,\bGamma)$ denote a $p$-dimensional stochastic process with vector of mean functions $\bfeta(t),t\in\calT$, and matrix  of covariance functions $\bGamma(s,t),s,t\in\calT$, where $\calT$ is the time period of interest, often a finite interval $[a,b]$ say with $ -\infty<a<b<\infty$.
Suppose we have $k$ independent functional samples
\begin{equation}\label{ksamp.sec2}
	\by_{i1}(t),\ldots,\by_{i n_i}(t)\iidsim \SP_p(\bfeta_i,\bGamma_i),\; i=1,\ldots,k,
\end{equation}
where $\bfeta_i(t),i=1,\ldots,k$ are the unknown vectors of group mean function of the $k$ samples, and $\bGamma_i(s,t)=\Cov[\by_{i1}(s),\by_{i1}(t)],i=1,\ldots,k$ are the unknown matrices of  covariance functions. Here $\bGamma_i$ is symmetric in the sense that $\bGamma_i(s,t)=\bGamma_i(t,s)^\top,i=1,\ldots,k$. Note that we assume the functional observations from the same group are independent and identically distributed (i.i.d.) and the functional observations from different groups are also independent.  One may wish to test whether the $k$ mean functions are equal:
\begin{equation}\label{MANOVA.sec2}
	H_0:\bfeta_1(t)=\cdots =\bfeta_k(t), \;\forall t\in\calT,
\end{equation}
against the usual alternative that at least two of the vectors of mean function are not the same. For the above problem, several interesting tests have been proposed by a number of authors. First of all, when $p=1$, meaning that the FMANOVA problem in (\ref{MANOVA.sec2}) is simplified to one-way ANOVA problem for functional data (FANOVA). Numerous tests have been put forth for the FANOVA problem and have been employed in various practical scenarios,  such as ischemic heart screening and detecting changes in air pollution during the COVID-19 pandemic, including \cite{cuevas2004anova,cuesta2010simple,zhang2013analysis, zhang2014one,zhang2019new,acal2022functional}. In particular,  \citet[Chapter 6]{zhang2013analysis} proposed three methods for the GLHT problem for univariate functional data, namely, the pointwise $F$, $L^2$-norm-based, and $F$-type tests. \cite{smaga2019linear} studied the theoretical properties of the above $L^2$-norm-based and $F$-type tests,  and proposed two new testing procedures by globalizing the pointwise $F$-test through  taking the integral or the supremum over $\calT$. When $p>1$,  \cite{gorecki2017multivariate} firstly defined the ``between"
and ``within" matrices for MFD, and then used them to construct several test statistics, including
the Wilks lambda test statistic, the Lawley--Hotelling trace test statistic, the Pillai trace test statistic and Roy's
maximum root test statistic. \cite{Qiu2021} then proposed two global tests for the two-sample problem for MFD, and recently, \cite{qiu2024tests} extended this work to the multi-sample problem. \cite{Zhu2021} studied the Lawley--Hotelling trace test for the FMANOVA problem by assuming that the matrices of covariance functions of the $k$ samples are the same, while \cite{zhu2023ksampFMD} introduced a global test statistic for the heteroscedastic FMANOVA problem. Although some work has been done for the one-way FMANOVA problem, the post hoc or contrast analysis is not satisfactorily developed in the literature. To the best of our knowledge, this study may be the first work for the above GLHT problem for MFD. 

In this study, given the $k$ independent samples (\ref{ksamp.sec2}), we are interested in testing the following GLHT problem for MFD:
\begin{equation}\label{GLHTH0.sec2}
	H_0: \bG\bM(t)=\bzero,\; t\in\calT \mbox{ vs. } H_1: \bG\bM(t)\neq\bzero, \; \mbox{for some }t\in\calT,
\end{equation}
where $\bM(t)=[\bm{\eta}_1(t),\ldots,\bm{\eta}_k(t)]^\top$ is a $k\times p$ matrix at time point $t$ whose rows are the $k$ mean functions, and $\bG: q\times k$ is a known coefficient matrix with $\operatorname{rank}(\bG)=q<k$. We assume that the matrix $\bG$ is of full row rank which implies that there are no redundant or contradictory hypotheses in (\ref{GLHTH0.sec2}). With properly specifying $\bG$, many particular hypothesis tests can be represented by a general testing framework described above. For example, when we set $\bG$ to be either  $(\bI_{k-1},-\bm{1}_{k-1})$ or $(-\bm{1}_{k-1},\bI_{k-1})$, a contrast matrix whose rows sum up to $0$, where $\bI_k$ and $\bm{1}_k$ denote the identity matrix of size $k\times k$ and a $k$-dimensional vector of ones, respectively, the GLHT problem (\ref{GLHTH0.sec2}) reduces to a one-way FMANOVA problem (\ref{MANOVA.sec2}). When the null hypothesis in (\ref{MANOVA.sec2}) is rejected, it is often of interest to further test if $\bfeta_1(t)=c_1\bfeta_2(t),\;\forall t\in\calT$ or if a contrast is zero, e.g., $c_2\bfeta_1(t)-(c_2+c_3)\bfeta_2(t)+c_3\bfeta_3(t)=\bzero, \;\forall t\in\calT$, where $c_1,c_2$, and $c_3$ are some known constants. To write the above two testing problems in the form of (\ref{GLHTH0.sec2}), we can set $\bG=(\bm{e}_{1,k}-c_1\bm{e}_{2,k})^\top$ and $\bG=(c_2\bm{e}_{1,k}-(c_2+c_3)\bm{e}_{2,k}+c_3\bm{e}_{3,k})^\top$, respectively, where and throughout $\bm{e}_{r,l}$ denotes a unit vector of length $l$ with the $r$-th entry being 1 and others 0. It is clear that the contrast matrix $\bG$ is not unique for the same hypothesis test. Therefore, it is important to construct a test which is invariant under the following transformation of the coefficient matrix:
\begin{equation}\label{trans.equ}
   \bG:\bG \to \bP\bG,
\end{equation}
where $\bP$ is any $q\times q$ non-singular matrix. In other words, any non-singular transformations of the coefficient matrix $\bm{G}$ yield the same hypothesis.  As a result,  it is reasonable to expect that the proposed test should exhibit invariance under any non-singular transformations of $\bm{G}$ (\ref{trans.equ}). To achieve this, we will rewrite the GLHT problem \eqref{GLHTH0.sec2} into the following equivalent form: 
\begin{equation}\label{GLHT1.sec2}
	H_0: \tbG\bM(t)=\bzero,\;t\in\calT \mbox{ vs. } H_1: \tbG\bM(t)\neq \bzero, \; \mbox{for some }t\in\calT,
\end{equation}
where $\tbG=(\bG\bD\bG^\top)^{-1/2}\bG$ and  
\mrk{
\begin{equation}\label{D.equ}
	\bD=\diag(1/n_1,\ldots,1/n_k). 
\end{equation}}
\mrk{Since  all the diagonal entries of $\bD$ are strictly positive, $\bG\bD\bG^\top$ is positive definite, and $(\bG\bD\bG^\top)^{-1}$ exists.}

In this article, we propose a new global test for the GLHT problem (\ref{GLHT1.sec2}) which can be employed in a wider range of scenarios or research contexts beyond the specific framework proposed by  \cite{gorecki2017multivariate,Qiu2021, Zhu2021,zhu2023ksampFMD}. Our main contributions can be described as follows. Firstly, we establish the pointwise variation matrices due to hypothesis for the GLHT problem (\ref{GLHT1.sec2}) and due to error, respectively, which serve as the foundation for constructing the pointwise test statistic. By globalizing the pointwise test statistic, we derive the test statistic for our new global test. Under some regularity conditions and null hypothesis, we demonstrate that the asymptotic distribution of the new global test statistic is a chi-squared-type mixture with the mixing coefficients dependent on the matrices of covariance functions.  It is worth highlighting that although the proposed global test for the GLHT problem in MFD shares similarities with the test statistic presented in \cite{zhu2023ksampFMD}, it is not simply an application of their results and there is no transformation of the GLHT problem into a standard heteroscedastic one-way FMANOVA problem considered in \cite{zhu2023ksampFMD}.
Secondly, since the asymptotic null distribution of the test statistic is a chi-squared-type mixture, instead of utilizing the Welch--Satterthwaite $\chi^2$-approximation as in \cite{zhu2023ksampFMD}, we employ the three-cumulant (3-c) matched $\chi^2$-approximation of \cite{zhang2005approximate} to approximate it with the parameters estimated from the data, which is expected to provide superior accuracy by considering not only the mean and variance of the test statistic but also its third moment.  Then, the proposed new global test can be conducted using the approximate critical value or the approximate $p$-value.  Thirdly, we introduce an adjustment coefficient that effectively enhances the performance of our proposed test, even with a relatively small sample size. Additionally,  we present fast computation methods as outlined in Section~\ref{NumImp} which has not been discussed in \cite{zhu2023ksampFMD}. Furthermore, we also derive asymptotic power of the new global test under some local alternatives, noting that the asymptotic properties of the proposed test statistic, as described in Theorems~\ref{T0.thm} and~\ref{T1.thm}, are not derived directly from the asymptotic properties of \cite{zhu2023ksampFMD}'s test for the standard heteroscedastic one-way FMANOVA problem. Finally, the simulation results presented in Section~\ref{simu.sec} demonstrate that our new global test has a wider range of applications and generally performs better or no worse than its competitors in terms of size control.

The rest of this paper is organized as follows. The main results are presented in Section~\ref{Main.sec3}. Simulation studies and real data applications are given in Sections~\ref{simu.sec} and~\ref{real.sec}, respectively.  Concluding remarks are provided  in Section~\ref{sec:conc}. Technical proofs for the main results are outlined in the Appendix. The R code used in this study is provided in the Supplementary Material.

\section{Main Results}\label{Main.sec3}
\subsection{Test Statistic}

For each $i=1,\ldots,k$, based on the $i$-th sample in (\ref{ksamp.sec2}) only,  the vector of mean functions $\bfeta_i(t)$ and matrix of covariance functions $\bGamma_i(s,t)$ can be unbiasedly estimated by
\begin{equation}\label{sampmeancovk}
\begin{array}{rcl}
   	\widehat{\bfeta}_i(t)&=&\barby_i(t)=n_i^{-1}\sum_{j=1}^{n_i}\by_{ij}(t),\;\mbox{ and }\;\\
    \hbGamma_i(s,t)&=&(n_i-1)^{-1}\sum_{j=1}^{n_i}[\by_{ij}(s)-\barby_i(s)][\by_{ij}(t)-\barby_i(t)]^{\top}, 
\end{array}
\end{equation}
which are known as vector of sample mean functions and matrix of sample covariance functions, respectively. For further study, we set 
\begin{equation}\label{xij.equ}
    \bx_{ij}(t)=\by_{ij}(t)-\bm{\eta}_i(t),j=1,\ldots,n_i;\;i=1,\ldots,k,
\end{equation}
and let $\barbx_i(t)$ be the vector of sample mean functions of $\bx_{ij}(t),j=1,\ldots,n_i;i=1,\ldots,k$, so that $\barbx_i(t)=\barby_i(t)-\bfeta_i(t),i=1,\ldots,k$.

To test (\ref{GLHT1.sec2}), we construct the pointwise  variation matrix due to hypothesis for the GLHT problem (\ref{GLHT1.sec2}) as 
\begin{equation}\label{Bn.equ}
    \bB_n(t)=[\bG\hbM(t)]^\top(\bG\bD\bG^\top)^{-1}[\bG\hbM(t)],
\end{equation}
where $\hbM(t)=[\barby_1(t),\ldots,\barby_k(t)]^\top$ is the usual unbiased estimator of $\bM(t)$] \mrk{and $\bD$ is defined in (\ref{D.equ})}. Let $\bbH=\bG^\top(\bG\bD\bG^\top)^{-1}\bG=(h_{ij}):k\times k$. It follows that 
    $\bB_n(t)=\hbM(t)^\top\bbH\hbM(t) = \sum_{i=1}^k\sum_{j=1}^k h_{ij}\barby_i(t)\barby_j(t)^\top$.
Under  $H_0$  in (\ref{GLHT1.sec2}), we can further express $\bB_n(t)$ as 
    $\bB_{n,0}(t)=[\hbM(t)-\bM(t)]^\top\bbH[\hbM(t)-\bM(t)]=\sum_{i=1}^k\sum_{j=1}^kh_{ij}\barbx_i(t)\barbx_{j}(t)^\top$.
 Since the $k$ functional samples (\ref{ksamp.sec2}) are independent, the expectation of $\bB_n(t)$ under $H_0$ is
$\bOmega_n(t,t)=\E[\bB_{n,0}(t)]=
	\sum_{i=1}^kh_{ii}\bGamma_i(t,t)/n_i$.
It is clear that the unbiased estimator of $\bOmega_n(t,t)$ is 
\begin{equation}\label{omegan.equ}
    \hbOmega_n(t,t)=\sum_{i=1}^kh_{ii}\hbGamma_i(t,t)/n_i,
\end{equation}
where the sample matrix of covariance functions $\hbGamma_i(t,t), i=1,\ldots,k$ is given in (\ref{sampmeancovk}). We can regard $\hbOmega_n(t,t)$ as the pointwise variation matrix due to error. 

At each time point $t\in\calT$, we can construct the following test statistic:
\begin{equation}\label{Tnt1.sec2}
	T_n(t)=\tr[\bB_n(t)\hbOmega_n^{-1}(t,t)],
\end{equation}	
where $\operatorname{tr}(\bm{A})$ denotes  the trace of a square matrix $\bm{A}$. Throughout this paper, we assume that the dimension $p$ is fixed and it is smaller than the total sample size $n=n_1+\cdots+n_k$. Consequently, $\hbOmega_n(t,t)$ is invertible almost surely at each time point $t\in\calT$.  However, even if the pointwise test is significant for all $t\in\calT$ at a given significance level, it does not ensure that the alternative hypothesis is overall significant at the same level. Additionally, conducting the pointwise test at all $t\in\calT$ can be time-consuming. To overcome these difficulties, we propose a new global test via globalizing the pointwise  test statistic $T_n(t)$ in (\ref{Tnt1.sec2}) using its integral over $\calT$:
\begin{equation}\label{teststa.sec3}
	T_{n}=\int_{\calT}T_n(t)dt=\int_{\calT}\tr[\bB_n(t)\hbOmega_n^{-1}(t,t)]dt.
\end{equation}
We reject the null hypothesis (\ref{GLHT1.sec2}) for large values of $T_n$. As a result,  our objective in the next section is to explore the null distribution of $T_{n}$ as defined in (\ref{teststa.sec3}).
\subsection{Asymptotic Null Distribution}
Throughout this paper,  let $\LpT$ denote the Hilbert space of $p$-dimensional vectors of square integrable functions on the interval $\calT$, that is, $\|\bm{f}\|_{\HS}=\left(\int_{\calT}||\bm{f}(t)||^2dt\right)^{1/2}<\infty, \;\bm{f}(t)\in \LpT$,  where $\|\cdot\|_{\HS}$ denotes the Hilbert--Schmidt norm and $\|\cdot\|$ denotes the usual $L^2$-norm of a $p$-dimensional vector in $\mathbb{R}^p$. Then the associated inner-product is defined as: $<\bm{f},\bm{g}>_{\HS}=\int_{\calT}\bm{f}(t)^\top\bm{g}(t)dt,\; \bm{f}(t),\bm{g}(t)\in \LpT$. See details in \cite{gorecki2018selected}. Let $\|\bA\|_{\F}$ denote the Frobenius norm of a matrix $\bA\in\mathbb{R}^{m\times n}$ with $\|\bA\|_{\F}=(\sum_{u=1}^m\sum_{v=1}^na_{uv}^2)^{1/2}$. We use $\calT^2$ and $\calT^3$  to denote the direct products of $\calT\times\calT$ and  $\calT \times \calT \times \calT$, respectively.  For our theoretical study, we need the following conditions:
\begin{enumerate}[C1.]
	\item[C1.] For each $i=1,\ldots,k$, 
	assume $\bm{\eta}_i(t)\in {\LpT}$ and $\int_{\calT^2}\|\bGamma_i(s,t)\|_{\F}^2dsdt<\infty$.
	\item[C2.] Let $n_{\min}=\min_{i=1}^kn_i$. As $n_{\min}\to\infty$, the $k$ group sizes satisfy $n_i/n\to\tau_i,i=1,\ldots,k$, for some constants $\tau_1,\ldots,\tau_k\in(0,1)$.
	\item[C3.] The vectors of subject-effect functions $\bx_{ij}(t),j=1,\ldots,n_i;\;i=1,\ldots,k$ in (\ref{xij.equ}) are i.i.d..
	\item[C4.] For each $i=1,\ldots,k$, the vector of subject-effect functions $\bx_{i1}(t)$ satisfies $\E\|\bx_{i1}\|_{\HS}^4<\infty$, and for any $s,t\in\calT$, the expectation $\E(\|\bx_{i1}(s)\|^2\|\bx_{i1}(t)\|^2)$ is uniformly bounded. 
\end{enumerate}
Condition C1 is regular. When the $k$ functional samples (\ref{ksamp.sec2}) are Gaussian, it is easy to show that under Condition C1, the vector of sample mean functions $\barby_i(t)$ is a $p$-dimensional Gaussian process and the matrix of sample covariance functions $\hbGamma_i(s,t)$ is proportional to a $p\times p$ dimensional Wishart process, for $i=1,\ldots,k$. Condition C2 requires that the $k$ sample sizes $n_1,\ldots,n_k$ tend to $\infty$ proportionally. When the $k$ functional samples (\ref{ksamp.sec2}) are non-Gaussian,  Conditions {C1--C3} guarantee that as $n_{\min}\to\infty$, the vector of sample group mean functions $\barby_i(t)$ converges weakly to a $p$-dimensional Gaussian process, $i=1,\ldots,k$. 
Condition C4 is used to ensure the pointwise convergence $\hbGamma_i(t,t)\convP \bGamma_i(t,t)$ for all $t\in\calT, i=1,\ldots,k$, as $n_{\min}\to\infty$, where here and throughout $\convP$ means convergence in probability. This condition is similar to Condition C4 in \cite{qiu2024tests}. By Lemma A.1 of  \cite{qiu2024tests}, we have $\hbOmega_n(t,t)\convP\bOmega_n(t,t)$ uniformly over $\calT$ when $n_{\min}\to\infty$. Consequently, using the continuous mapping theorem, $\hbOmega_n^{-1}(t,t)\convP\bOmega_n^{-1}(t,t)$ uniformly holds as well. 



By setting $T_n^*=\int_{\calT}\tr[\bB_n(t)\bOmega_n^{-1}(t,t)]dt$, we see that $T_n$ and $T_n^*$ have the same distribution for large values of $n_{\min}$, and therefore,
studying the asymptotic null distribution of $T_n$ is equivalent to studying that of $T_n^*$. For further discussion, let $\bX(t)=[\barbx_1(t),\ldots,\barbx_k(t)]^\top$, we now express $ T^*_{n}$ as follows:
\begin{equation}\label{Tnde.eqn}
    T^*_{n}  = T^*_{n,0} + 2S^*_n+\int_{\calT}\tr[\bM(t)^\top\bbH\bM(t)\bOmega_n^{-1}(t,t)]dt,
\end{equation}
where
\mrk{
\begin{equation}\label{Tnde.sec2}
	\begin{split}
			T^*_{n,0}  &= \int_{\calT}\tr\left[\bX(t)^\top\bbH\bX(t)\bOmega_n^{-1}(t,t)\right]dt,\;\mbox{ and }\; \\
		S^*_{n} &= \int_{\calT}\tr\left[\bX(t)^\top\bbH\bM(t)\bOmega_n^{-1}(t,t)\right]dt. 
	\end{split}
   \end{equation}}
It can be inferred that $T^*_{n,0}$ has the same distribution as that of $T^*_n$ under the null hypothesis. \mrk{Note that under the transformation in  (\ref{trans.equ}),   we have  $(\bP\bG\bD\bG^{\top}\bP^{\top})^{-1}=(\bP^{-1})^\top(\bG\bD\bG^\top)^{-1}\bP^{-1}$. It follows that $\bbH\to \bG^\top\bP^\top(\bP^{-1})^\top(\bG\bD\bG^\top)^{-1}\bP^{-1}\bP\bG=\bbH$, which remains invariant under the non-singular transformation (\ref{trans.equ}). Therefore,  the invariance of $\bOmega_n(t,t)$, $\hbOmega_n(t,t)$, $T^*_{n,0}$, $S^*_n$, and $\int_{\calT}\tr[\bM(t)^\top\bbH\bM(t)\bOmega_n^{-1}(t,t)]dt$ follows immediately. }


Before we derive the asymptotic distribution of $T^*_{n,0}$  in (\ref{Tnde.sec2}), we need the following useful notations. For simplicity of notation,  throughout this paper, for any covariance function matrix $\bGamma_i(s,t), s,t\in\calT,i=1,\ldots,k$, we write $\tr(\bGamma_i)=\int_{\calT}\tr\left[\bGamma_i(t,t)\right]dt,  \tr(\bGamma_i^{2})=\int_{\calT^2}\tr\left[\bGamma_i(s,t)\bGamma_i(t,s)\right]dsdt$, and $\tr(\bGamma_i^{3})=\int_{\calT^3}\tr\left[\bGamma_i(s,t)\bGamma_i(t,v)\bGamma_i(v,s)\right]dsdtdv$; for any two distinct covariance function matrices $\bGamma_i(s,t)$ and $\bGamma_{j}(s,t),s,t\in\calT,\;i\neq j$, we write  $\tr(\bGamma_i\bGamma_{j})=\int_{\calT^2}\tr\left[\bGamma_i(s,t)\bGamma_{j}(t,s)\right]dsdt$;  and for any three distinct covariance function matrices $\bGamma_i(s,t), \bGamma_j(s,t), \bGamma_\ell(s,t), s,t\in\calT, i\neq l\neq \ell$,  we write
$\tr(\bGamma_i\bGamma_j \bGamma_\ell)=\int_{\calT^3}\tr\left[\bGamma_i(s,t)\bGamma_j(t,v)\bGamma_\ell(v,s)\right]dsdtdv$. Furthermore, we use  $\chi_v^2$ to denote a central chi-squared distribution with $v$ degrees of freedom, $\convL$ to denote convergence in distribution, $\convd$ to denote equality in distribution, and $\otimes$ to denote the Kronecker product \mrk{ \citep[see, e.g.,][]{horn1994topics}}.  The following theorem shows that the asymptotic distribution of $T_{n,0}^*$ is a central $\chi^2$-type mixture. The proof of Theorem~\ref{T0.thm} is given in the Appendix.

\begin{theorem}\label{T0.thm} 	Assume Conditions C1--C3 hold. Then, as $n_{\min}\to\infty$, we have
	$T_{n,0}^*\convL T_0^*$, where $T_0^*\convd  \sum_{r=1}^\infty \lambda_rA_r,$
	with	$A_1,A_2,\ldots \iidsim \chi_1^2$ and $\lambda_1,\lambda_2,\ldots$ being the  eigenvalues of $\bSigma(s,t)=\bC\diag[\bGamma_1^*(s,t),\ldots,\bGamma_k^*(s,t)]\bC^\top$ in descending order with  
    \begin{equation}\label{Gammastar.equ}
        \bGamma^*_i(s,t)=\bOmega_n^{-1/2}(s,s)\bGamma_i(s,t)\bOmega_n^{-1/2}(t,t)/n_i, i=1,\ldots,k,
    \end{equation}
    and $\bC=[(\bG\bD\bG^\top)^{-1/2}\bG]\otimes \bI_p$. In addition, the first three cumulants of $T_0^*$ are given by 
    \begin{equation}\label{3c2.sec2}
\begin{array}{rcl}
	\calK_1(T_0^*)&
    =&(b-a)p,\;
	\calK_2(T_0^*)
	=2\sum_{i=1}^k\sum_{j=1}^kh_{ij}^2\tr(\bGamma_i^*\bGamma_{j}^*),\mbox{ and }\\
	\calK_3(T_0^*)&=&8\sum_{i=1}^k\sum_{j=1}^k\sum_{\ell=1}^kh_{ij}h_{j\ell}h_{\ell i}\tr(\bGamma_i^*\bGamma_{j}^* \bGamma_{\ell}^*).
\end{array}
\end{equation}
\end{theorem}

\subsection{Approximations to the Null Distribution}

Theorem~\ref{T0.thm} indicates that the null distribution of our test statistic $T_{n}$ is asymptotically a $\chi^2$-type mixture $T_0^*$, which can be approximated by the three-cumulant (3-c) matched $\chi^2$-approximation  introduced by  \cite{zhang2005approximate}. The fundamental idea behind the 3-c matched $\chi^2$-approximation is to approximate the distribution of $T_0^*$ by that of a random variable $R$ represented as 
$R\dequ \beta_0+\beta_1\chi_d^2$ via matching their first three  cumulants, namely, means, variances, and third central moments. The specific values of $\beta_0$, $\beta_1$, and $d$ are determined by equating the means, variances, and third central moments of $T_0^*$ and $R$. In comparison to  the normal approximation and the well-known Welch--Satterthwaite $\chi^2$-approximation (\citealt{welch1947generalization,satterthwaite1946approximate}), both of which  are two-cumulant matched approximation, the 3-c matched  $\chi^2$-approximation is anticipated to offer superior accuracy since it not only matches the mean and variance of the test statistic, but also takes the third moment of the test statistic into account.

 The first three cumulants of $R$ are given by $\calK_1(R)=\beta_0+\beta_1d$, $\calK_2(R)=2\beta_1^2d$, and $\calK_3(R)=8\beta_1^3d$, while the first three cumulants of  $T_0^*$ are given in Theorem~\ref{T0.thm}. Therefore, we have
\[
	\beta_0=\calK_1(T_0^*)-\frac{2\calK_2^2(T_{0}^*)}{\calK_3(T_{0}^*)},\;\beta_1=\frac{\calK_3(T_{0}^*)}{4\calK_2(T_{0}^*)},\;\mbox{ and }\;d=\frac{8\calK_2^3(T_{0}^*)}{\calK_3^2(T_{0}^*)}.
\]

The proposed test can be implemented provided that the parameters $\beta_0$, $\beta_1$, and $d$ are properly
estimated. 
To apply the 3-c matched $\chi^2$-approximation, we need to estimate  $\calK_2(T_0^*)$ and $\calK_3(T_0^*)$ in (\ref{3c2.sec2}) properly. 
Based on the given $k$ functional samples (\ref{ksamp.sec2}), we can obtain the
following naive estimators of  $\calK_2(T_0^*)$ and $\calK_3(T_0^*)$ by replacing
$\bGamma_i^*(s,t)$ in (\ref{3c2.sec2}) with their estimators $\hbGamma_i^*(s,t)=\hbOmega^{-1/2}_n(s,s)\hbGamma_i(s,t)\hbOmega^{-1/2}_{n}(t,t)/n_i,i=1,\ldots,k$:
	\[
 \begin{split}
     \widehat{\calK_2(T_0^*)}= 2\sum_{i=1}^k\sum_{j=1}^kh_{ij}^2\tr(\hbGamma_i^*\hbGamma_j^*),\;\mbox{and }
     \widehat{\calK_3(T_0^*)}=8\sum_{i=1}^k\sum_{j=1}^k\sum_{\ell=1}^kh_{ij}h_{j\ell}h_{\ell i}\tr(\hbGamma_i^*\hbGamma_{j}^* \hbGamma_{\ell}^*).
 \end{split}
 \]
It follows that 
\begin{equation}\label{hatbetadN.sec3}
	\hbeta_0=(b-a)p-\frac{	2\widehat{\calK_2(T_0^*)}^2}{\widehat{\calK_3(T_0^*)}},\;\hbeta_1=\frac{\widehat{\calK_3(T_0^*)}}{4\widehat{\calK_2(T_0^*)}},\;\mbox{ and }\;\hd=\frac{8\widehat{\calK_2(T_0^*)}^3}{\widehat{\calK_3(T_0^*)}^2 }.
\end{equation}

\begin{remark}
This is a natural way to estimate the parameters $\beta_0$, $\beta_1$, and $d$, however, the estimators in (\ref{hatbetadN.sec3}) are biased. To reduce the bias, \cite{zhu2023ksampFMD} also proposed a bias-reduced method to estimate their parameters. Based on their simulation studies, they found that when the within-subject observations are highly or moderately correlated, the bias-reduced method is more liberal than naive method, and  when the within-subject observations are less correlated, the bias effect may not be ignorable and naive method is more conservative than bias-reduced method. Therefore, they recommended using the naive method when the within-subject observations are highly or moderately correlated, and  using the bias-reduced method when the within-subject observations are less correlated. Nonetheless, when conducting real data analysis, it is challenge to assess whether the within-subject observations are less correlated, moderately correlated, or highly correlated. Often, researchers may resort to employing the naive method or bias-reduced method without a precise evaluation of the correlation structure due to practical limitations or complexities.  In addition,  the bias-reduced estimators are calculated when $n_{\min}$ is large. However, when dealing with smaller sample sizes, this bias-reduced method tends to exhibit a liberal bias. This is actually confirmed by the simulation results presented in Tables~\ref{size1.tab}, \ref{size4.tab}, and~\ref{size5.tab}  of Section~\ref{simu.sec}.
\end{remark}

In this paper, rather than suggesting a complex and time-consuming bias-reduced approach, we follow the ideas of \cite{srivastava2008test,zhang2023two,cao2024scale} and introduce an adjustment coefficient that effectively enhances the performance of our proposed test across varying levels of within-subject correlation, even with a relatively small sample size. For each $i=1,\ldots,k$, by Lemma 2 of \cite{zhu2023ksampFMD} and when $n_{\min}$ is large, the bias-reduced estimator of $\calK_2(T_0^*)$ is given by
\begin{equation*}\label{k2b.equ}
\begin{array}{rcl}
     \widehat{\calK_2(T_0^*)}_{B} &=&  \widehat{\calK_2(T_0^*)}\\
     &+& 2\sum_{i=1}^{k}\frac{(n_i+1)h_{ii}^2}{n_i(n_i-3)}\tr(\hbGamma_{i}^{* 2})+ 2 \sum_{i=1}^{k}\frac{(n_i-1)h_{ii}^2}{n_i(n_i-2)(n_i-3)}\Big[ \tr^2(\hbGamma_i^*)-n_i Q^*_i\Big],
\end{array}
	\end{equation*}
where  $Q_{i}^*=\sum_{i=1}^{n_i}\|\hbOmega_n^{-1/2}(\by_{ij}-\barby_i)\|_{\HS}^4/[n_i^2(n_i-1)],i=1,\ldots,k$. Incorporating the term involving $\tr(\hbGamma_{i}^{* 2})$,  we propose the following adjustment coefficient $c_n$:
\begin{equation}\label{cn.equ}
    c_n=1+\sum_{i=1}^k h_{ii}^2\frac{n_i+1}{n_i(n_i-3)}\tr(\hbGamma_{i}^{* 2}).
\end{equation}
It is seen that the adjustment coefficient $c_n\to 1$ as $n_{\min}\to\infty$. 
Therefore, for any nominal significance level $\alpha>0$, let $\chi_v^2(\alpha)$ denote the upper $100\alpha$ percentile of the $\chi_v^2$ distribution. The proposed test is conducted by computing the $p$-value using the following approximate null distribution:  $T_n/c_n\sim \hbeta_0+\hbeta_1\chi_{\hd}^2\,\mbox{ approximately under } H_0$,
or we reject the null hypothesis $H_0$ whenever $ T_n/c_n> \hbeta_0+\hbeta_1\chi_{\hd}^2(\alpha)$, where $\hbeta_0$, $\hbeta_1$, and $\hd$ are the naive estimators in (\ref{hatbetadN.sec3}).

\subsection{Asymptotic Power}

In this section, we investigate the asymptotic power of our new global test, based on the test statistic $T_n$ given in (\ref{teststa.sec3}) and the 3-c matched $\chi^2$-approximation. It is evident from the expression of $T_n^*$ in (\ref{Tnde.eqn}) that  $T^*_{n,0}$ has the same distribution as $T_n^*$ under $H_0$. In addition, $S_n^*$  in (\ref{Tnde.sec2}) can be further expressed as $S_n^*=\int_{\calT}\sum_{i=1}^k\sum_{j=1}^kh_{ij}\barbx_i(t)^\top\bOmega_n^{-1}(t,t)\bfeta_j(t)dt=\int_{\calT}\barbx^*(t)^\top(\bbH\otimes\bI_p)\bm{\eta}^*(t)dt$, where $\barbx^*(t)=[\barbx_1(t)^\top\bOmega_n^{-1/2}(t,t),\ldots,\barbx_k(t)^\top\bOmega_n^{-1/2}(t,t)]^\top$ and  $\bm{\eta}^*(t)=[\bfeta_1(t)^\top\bOmega_n^{-1/2}(t,t),\ldots,\bfeta_k(t)^\top\bOmega_n^{-1/2}(t,t)]^\top$ are two long vectors of dimension $kp$. It is also noteworthy  that $\E(S_n^*)=0$ and
\[
\begin{array}{rcl}
      \Var(S_n^*)&=&\E({S_n^*}^2)=\int_{\calT^2}\bm{\eta}^*(s)^\top(\bbH\otimes\bI_p)\E[\barbx^*(s)\barbx^*(t)^\top](\bbH\otimes\bI_p)\bm{\eta}^*(t)dsdt\\
      &=&\int_{\calT^2}\bm{\eta}^*(s)^\top\bC^\top\bC\diag[\bGamma_1^*(s,t),\ldots,\bGamma_k^*(s,t)]\bC^\top\bC\bm{\eta}^*(t)dsdt\\
      &=&\int_{\calT^2}\bm{\eta}^*(s)^\top\bC^\top\bSigma(s,t)\bC\bm{\eta}^*(t)dsdt,\\
\end{array}
\]
where $\bSigma(s,t)=\bC\diag[\bGamma_1^*(s,t),\ldots,\bGamma_k^*(s,t)]\bC^\top$ as defined in Theorem~\ref{T0.thm}.

For simplicity, we investigate the asymptotic power of $T_{n}$ under the following specified local alternative:
\begin{equation}\label{H1.sec3}
	\Var(S_n^*) = o[\tr(\bSigma^{2})].
\end{equation}
This condition describes the case when the information in the local alternatives is relatively small compared with the variance of $T^*_{n,0}$. It allows that the test statistic $T_{n}$ is mainly dominated by $T^*_{n,0}$ since under the local alternative (\ref{H1.sec3}), we have $S_{n}^*/\tr(\bSigma^{ 2})\convP 0$.  
\begin{remark}
    The local alternative (\ref{H1.sec3}) has been widely used in equal-mean testing problem for high-dimensional data. For instance, when $k=2$, the GLHT problem in (\ref{GLHT1.sec2}) reduces to the two-sample equal-mean function testing problem. It follows that  $\bSigma(s,t)=n^{-1}n_1n_2[\bGamma_1^*(s,t)+\bGamma_2^*(s,t)]$, and $\Var(S_n^*)=n^{-1}n_1n_2\int_{\calT^2}[\bfeta_1^*(s)-\bfeta_2^*(s)]^\top\bSigma(s,t)[\bfeta^*_1(t)-\bfeta^*_2(t)]dsdt$, where $\bGamma^*_i(s,t),i=1,\ldots,k$ is defined in (\ref{Gammastar.equ}).
  Thus, in this context, the  local alternative (\ref{H1.sec3}) is similar to those used in \cite{bai1996effect} and \cite{chen2010two}. Therefore, it  generalizes the conditions used in \cite{bai1996effect} and \cite{chen2010two}   in the context of MFD and is similar to Equation (21) in \cite{zhang2022testing}, Equation (23) in \cite{zhu2022linear}, Condition C7 in \cite{cao2024scale}, and Assumption D in  \cite{li2024linear}, respectively.
\end{remark} 
In particular, under Condition C2, as $n_{\min}\to\infty$, we have 
\mrk{
\begin{equation}\label{limitOmegan.sec2}
\begin{array}{rcl}
    	&&n^{-1}\bbH \to \bbH^* =\bG^\top(\bG\bD^*\bG^\top)^{-1}\bG,\; \\
    	&&\bOmega_n(t,t)\to \bOmega(t,t)=\sum_{i=1}^k h^*_{ii}\bGamma_i(t,t)/\tau_i,	 \mbox{ and }\\
	 &&\tr(\bSigma^{2})\to 	\sum_{i=1}^k\sum_{j=1}^kh_{ij}^{*2}\tr(\tilde{\bGamma}^*_{i}\tilde{\bGamma}^*_{j}),
\end{array}
\end{equation}}
where $\bD^*=\diag(1/\tau_1,\ldots,1/\tau_k)$,   $\tilde{\bGamma}^*_{i}(s,t)=\bOmega^{-1/2}(s,s)\bGamma_{i}(s,t)\bOmega^{-1/2}(t,t)/\tau_i,i=1,\ldots,k$, and $h_{ij}^*$ is the $(i,j)$-entry of $\bbH^*$. We then have the following theorem. Proof of Theorem~\ref{T1.thm} is given in the Appendix.
\begin{theorem}\label{T1.thm}
	Assume Conditions {C1--C4} hold, and $\hbeta_0$, $\hbeta_1$, and $\hd$ are ratio-consistent for $\beta_0$, $\beta_1$, and $d$, respectively.  Then under the local alternative (\ref{H1.sec3}) and as $n_{\min}\to\infty$, we have
	\[
 \begin{split}
     &\quad\Pr\left[T_{n}/c_n\geq \hbeta_0+\hbeta_1\chi_{\hd}^2(\alpha)\right]\\
     &=\Pr\left\{\frac{\chi_{d}^2-d}{\sqrt{2d}}\geq \frac{\chi_{d}^2(\alpha)-d}{\sqrt{2d}}-\frac{n\int_{\calT}\tr\left[\bM(t)^\top\bbH^*\bM(t)\bOmega^{-1}(t,t)\right]dt}{\sqrt{2	\sum_{i=1}^k\sum_{j=1}^kh_{ij}^{*2}\tr(\tilde{\bGamma}^*_{i} \tilde{\bGamma}^*_{j})}}\right\}[1+o(1)].
 \end{split}		
		\]
	where 
    $\bbH^*, \bOmega(t,t),t\in\calT$, and $\tilde{\bGamma}^*_{i}(s,t),i=1,\ldots,k$ are defined in (\ref{limitOmegan.sec2}).
\end{theorem}
\begin{remark}
In particular, if $d\to\infty$,  we have $(\chi_{d}^2-d)/\sqrt{2d}\to N(0,1)$ and  $[\chi_{d}^2(\alpha)-d]/\sqrt{2d}\to z_{\alpha}$, where $z_{\alpha}$ denotes the upper $100\alpha$-percentile of $N(0,1)$. Thus, the asymptotic power of $T_n$ in Theorem~\ref{T1.thm} can be written as
\[
     \Pr\left[T_{n}/c_n\geq \hbeta_0+\hbeta_1\chi_{\hd}^2(\alpha)\right]=\Phi\left\{-z_{\alpha}+\frac{n\int_{\calT}\tr\left[\bM(t)^\top\bbH^*\bM(t)\bOmega^{-1}(t,t)\right]dt}{\sqrt{2	\sum_{i=1}^k\sum_{j=1}^kh_{ij}^{*2}\tr(\tilde{\bGamma}^*_{i} \tilde{\bGamma}^*_{j})}}\right\}[1+o(1)].
\]
Theorem~\ref{T1.thm} indicates that the asymptotic power of $T_n$ is mainly determined by the following ratio: 
\begin{equation}\label{ratio.eqn}
    \frac{n\int_{\calT}\tr\left[\bM(t)^\top\bbH^*\bM(t)\bOmega^{-1}(t,t)\right]dt}{\sqrt{2	\sum_{i=1}^k\sum_{j=1}^kh_{ij}^{*2}\tr(\tilde{\bGamma}^*_{i} \tilde{\bGamma}^*_{j})}}.
\end{equation}
As $n_{\min}\to\infty$, it is seen that the numerator in (\ref{ratio.eqn}) is the measure of the departure of the local alternative from the null hypothesis and the denominator in (\ref{ratio.eqn}) is the standard deviation of the null distribution.  The asymptotic power will be non-trivial if this ratio is positive and will tend to 1 if this ratio tends to $\infty$.
\end{remark}
\subsection{Numerical Implementation}\label{NumImp}
In the previous sections, the multivariate functional observations are assumed to be observed continuously, which results in easier presentation of the theory.  However, in practical situations, the $k$ functional samples (\ref{ksamp.sec2}) may not be observed continuously but at design time points, and these observation points may vary among different curves. When all the components of all the individual functional observations are observed at a common grid of design time points, we can apply the proposed test $T_n$ directly. When the design time points are not the same, we can use some existing smoothing techniques to smooth the curves first, and then discretize each component of the reconstructed functional observations at a common grid of time points, see details in \cite{zhang2014one}. This process essentially aligns with the ``smoothing first, then estimation" method investigated in \cite{zhang2007statistical} which demonstrates that under certain mild conditions, the asymptotic impact of the substitution effect can be disregarded.

 Now we assume all the components of all the individual functional observations are discretized at a common grid of resolution time points.  Let the resolution be $M$, a large number and let $t_1,\ldots,t_M$ be $M$ resolution time points which are equally spaced in $\calT$. 	Let $\bz_{ij}(t)=\by_{ij}(t)-\barby_i(t),j=1,\ldots,n_i;\;i=1,\ldots,k$. Then the matrix of sample covariance function $\hbGamma_i(s,t)$ in (\ref{sampmeancovk}) can be written as $\hbGamma_i(s,t)=(n_i-1)^{-1}\sum_{j=1}^{n_i}\bz_{ij}(s)\bz_{ij}(t)^\top $ and then be discretized  as
$\hbGamma_i(t_m,t_{m'})=(n_i-1)^{-1}\sum_{j=1}^{n_i}\bz_{ij}(t_m)\bz_{ij}(t_{m'})^\top$ accordingly. So are $\bB_n(t)$ (\ref{Bn.equ}) and  $\hbOmega_n(s,t)$ (\ref{omegan.equ}).

Let $v(\calT)$ denote the volume of $\calT$. When $\calT=[a,b]$, we have $v(\calT)=b-a$. It is sufficient to replace integrals by summations, then the test statistic can be discretized as
$	T_n\approx [v(\calT)/M]\sum_{m=1}^M\tr[\bB_n(t_m)\hbOmega_n^{-1}(t_m,t_m)]$,
where $\bB_n(t_m)=\hbM(t_m)^\top\bbH\hbM(t_m)$ and $\hbOmega_n^{-1}(t_m,t_m)=\sum_{i=1}^kh_{ii}\hbGamma_i(t_m,t_{m})/n_i$.
Before estimating $\beta_0$, $\beta_1$, and $d$ by the 3-c matched $\chi^2$-approximation, we can rewrite $\tr(\hbGamma_i^*)$, $\tr(\hbGamma_i^* \hbGamma_j^*)$ and $\tr(\hbGamma_i^*\hbGamma_j^* \hbGamma_\ell^*)$ as 
\[
\begin{split}
    &\tr(\hbGamma_i^*)=\frac{\sum_{u=1}^{n_i}\delta_{uu}^{ii}}{n_{i}(n_i-1)}, \; \tr(\hbGamma_i^*\hbGamma_j^*)=\frac{\sum_{u=1}^{n_i}\sum_{v=1}^{n_j}(\delta_{uv}^{ij})^2}{n_i n_j(n_i-1)(n_j-1)},\;\mbox{ and }\;\\
    &\tr(\hbGamma_i^* \hbGamma_j^*\hbGamma_\ell^*)=\frac{\sum_{u=1}^{n_i}\sum_{v=1}^{n_j}\sum_{s=1}^{n_{\ell}}\delta_{us}^{i\ell}\delta_{uv}^{ij}\delta_{vs}^{j\ell}}{n_i n_j n_\ell(n_i-1)(n_j-1)(n_{\ell}-1)},
\end{split}
\]
respectively, where $\delta_{uv}^{ij}=\int_{\calT}\bz_{iu}(t)^\top\hbOmega_n^{-1}(t,t)\bz_{jv}(t)dt,\;u=1,\ldots,n_i;\;v=1,\ldots,n_j;\;i,j=1,\ldots,k$.
Let $\bDelta_{ij}= (\delta_{uv}^{ij}):n_{i}\times n_{j},i,j=1,\ldots,k$. For fast computation, we have
\[
\begin{split}
   	&\tr(\hbGamma_i^*)=\frac{\tr(\bDelta_{ii})}{n_i(n_i-1)},\;	\tr(\hbGamma_i^* \hbGamma_j^*)=\frac{\tr(\bDelta_{ij}\bDelta_{ij}^\top)}{n_i n_j(n_i-1)(n_j-1)},\;\mbox{ and }\;\\
    &\tr(\hbGamma_i^* \hbGamma_j^*\hbGamma_\ell^*)=\frac{\tr(\bDelta_{ij}\bDelta_{j\ell}\bDelta_{\ell i})}{n_i n_j n_\ell(n_i-1)(n_j-1)(n_{\ell}-1)}.
\end{split}
\]

Since the number of groups, $k$, is fixed and usually not large, the above calculations are not time-consuming.  The values of $\delta_{uv}^{ij},u=1,\ldots,n_i;\;v=1,\ldots,n_j;\;i,j=1,\ldots,k$ can be discretized as
\begin{equation}\label{delta.equ}
\delta_{uv}^{ij}\approx \frac{v(\calT)}{M}\sum_{m=1}^M\bz_{iu}(t_m)^\top\hbOmega_n^{-1}(t_m,t_m)\bz_{jv}(t_m),
\end{equation}
so are $\bDelta_{ij},i,j=1,\ldots,k$. Then the estimators of $\beta_0$, $\beta_1$, and $d$ in (\ref{hatbetadN.sec3}) can be discretized based on the values of $\delta_{uv}^{ij},u=1,\ldots,n_i;\;v=1,\ldots,n_j;\;i,j=1,\ldots,k$  in (\ref{delta.equ}).

\section{Simulation Studies}\label{simu.sec}
In this section, we investigate the finite-sample behavior of the proposed test, denoted as $T_{\NEW}$, in various hypothesis testing problems  including the two-sample problem, one-way FMANOVA, and some specific linear hypotheses, and compare it to several established methods used for MFD. In addition, we also demonstrate the use of  $T_{\NEW}$ not only for dense functional data but also for sparse ones.  We compute the empirical size or power of a test as the proportion of the number of rejections out of $1000$ simulation runs.  Throughout this section, we set the nominal size as $\alpha=5\%$.

 \subsection{Simulation 1}\label{simu1.sec}
 
In this simulation, we consider  $k=3$ groups of multivariate functional samples with four vectors $\bn=[n_1,n_2,n_3]$ of sample sizes  $\bn_1=[7,12,12]$,  $\bn_2=[15,21,21]$, $\bn_3=[30,42,42]$, and $\bn_4=[50,70,70]$. 
 The multivariate functional samples are generated using the model: $\by_{ij}(t)=\bfeta_i(t)+\sum_{r=1}^q\sqrt{\lambda_{ir}}\varepsilon_{ijr}\bphi_r(t)$, $j=1,\ldots,n_i; \;i=1,\ldots,3$, where $\varepsilon_{ijr}$ are i.i.d. random variables, and the vector of orthonormal basis functions $\bm{\phi}_r(t)$, along with the variance components $\lambda_{ir},r=1,\ldots,q$ in descending order,  are used to flexibly specify the matrix of covariance functions $\bGamma_i(s,t)=\sum_{r=1}^q\lambda_{ir}\bm{\phi}_r(s)\bm{\phi}_r(t)^{\top}$, $i=1,\ldots,3$. Since the performance of the test is  not significantly affected by  the number of resolution time points $M$, as demonstrated in studies such as \cite{qiu2024tests} and \cite{munko2024multiple}, the generated functions are observed at $M=50$ equidistant points in the closed interval $\calT=[0,1]$. 
 We aim to compare the finite-sample performance of $T_{\NEW}$ and its competitors when the dimension $p$ of the generated MFD is reasonably large, and subsequently we set $p=6$.   The vector of mean functions $\bfeta_1(t)=[\eta_{11}(t),\ldots,\eta_{16}(t)]^\top$ for the first group is set as $\eta_{11}(t)=[\sin(2\pi t^2)]^5$, $\eta_{12}(t)=[\cos(2\pi t^2)]^5,\eta_{13}(t)=t^{1/3}(1-t)-5,\eta_{14}(t)=\sqrt{5}t^{2/3}\exp(-7t),\eta_{15}(t)=\sqrt{13t}\exp(-13t/2)$, and $\eta_{16}(t)=1+2.3t+3.4t^2+1.5t^3$.
To specify the other two group mean functions $\bfeta_2(t)$ and  $\bfeta_3(t)$, we set $\bfeta_2(t)=\bm{\eta}_1(t)+\delta \bm{g}(t)$ and  $\bfeta_3(t)=\bm{\eta}_1(t)+2\delta \bm{g}(t)$ in which $\delta$ controls the mean function differences $\bm{\eta}_i(t)-\bm{\eta}_1(t),i=2,3$, and $\bm{g}(t)$ controls the direction of these differences.  	For simplicity, we set the $\ell$-th entry of $\bm{g}(t)$ as $g_{\ell}(t)/[\sqrt{p}\|g_{\ell}\|]$ where $g_{\ell}(t)=(M-1)t^{\ell}+1,\ell=1,\ldots,p$.

 To specify the matrices of  covariance function $\bGamma_i(s,t)$, we set $\lambda_{i r}=\nu_{i}\rho^r,r=1,\ldots,q$, with $\rho=0.1,0.5$ and $0.9$ so that the components of the simulated functional data have high, moderate, or low correlations. Note that, for any fixed $i$, $\lambda_{ir},r=1,\ldots,q$, decays slowly if the value of $\rho$ is large, that is, the functional samples  become more noisy if the value of $\rho$ becomes larger. 
We set $\nu_1=1,\nu_2=2$, and $\nu_3=5$ to introduce heteroscedasticity among the three simulated functional sample.   For $\bm{\phi}_r(t)=[c_1\psi_r(t),\ldots,c_p\psi_r(t)]^\top$,   the basis functions are taken as $\psi_1(t)=1$, $\psi_{2r}(t)=\sqrt{2}\sin(2\pi rt)$, $\psi_{2r+1}(t)=\sqrt{2}\cos(2\pi rt),t\in\calT, r=1,\ldots,(q-1)/2$,  
and we let $c_{\ell}=\ell/(1^2+\cdots+p^2)^{1/2},\ell=1,\ldots,p$, so that $\sum_{\ell=1}^pc_{\ell}^2=1$ and 
$\|\bm{\phi}_r\|_{\HS}^2=1$ holds for $r=1,\ldots,q$. We take $q=7$.  To generate Gaussian functional data we set $\varepsilon_{ijr}\iidsim N(0,1)$, and we specify $\varepsilon_{ijr}\iidsim t_4/\sqrt{2}$
and $\varepsilon_{ijr}\iidsim (\chi_4^2-4)/(2\sqrt{2})$
to generate non-Gaussian functional data.

Now we consider the following hypothesis testing problems for MFD by employing  various coefficient matrices $\bG$:
\begin{enumerate}[H1.]
	\item[H1.]  Set $\bG$ to be $\bG_1=(\bI_2,-\bm{1}_2)$. Then the GLHT problem (\ref{GLHTH0.sec2}) reduces to the one-way FMANOVA problem (\ref{MANOVA.sec2}). We can examine the performance of $T_{\NEW}$ by comparing it to the permutation tests based on the Wilks', Lowley--Hotelling's, Pillay's, and Roy's test statistics from \cite{gorecki2017multivariate}, as well as the test proposed by \cite{zhu2023ksampFMD} with their naive and bias-reduced methods, denoted as  W, LH, P, R, $T_{\ZZC}^{\N}$, and $T_{\ZZC}^{\B}$, respectively,  in terms of size control and power.
	\item[H2.] Set $\bG$ to be  $\bG_2=(1,-1,0)$, $\bG_3=(1,0,-1)$, or $\bG_4=(0,1,-1)$. Then the GLHT problem (\ref{GLHTH0.sec2}) reduces to the two-sample problem which has been considered by \cite{Qiu2021}. Hence we can demonstrate the performance of $T_{\NEW}$ against the two global tests proposed by \cite{Qiu2021}, namely, the two-sample tests based on integrating and taking supremum of the pointwise Hotelling $T^2$-test statistic, denoted as $T_{\QCZ}$ and $T_{\QCZ}^{\max}$, respectively,  in terms of size control and power. 
	\item[H3.] Set $\bG$ to be $\bG_5=(1,-2,1)$. That is, we aim to test the following contrast test 
	\begin{equation}\label{H0G5.equ}
	H_0: \bfeta_1(t)-2\bfeta_2(t)+\bfeta_3(t)=\bzero, \; t\in\calT.
	\end{equation}
To examine the performance of $T_{\NEW}$, we compare it against a standard nonparametric bootstrap test, denoted as $T_{\BT}$. The steps of $T_{\BT}$ can be described as follows:
	\begin{enumerate}[(1)]
		\item[(1)]  Based on the given $k$ functional samples (\ref{ksamp.sec2}), compute the test statistics $T_n$ (\ref{teststa.sec3}) and record the residual functions $\bz_{ij}(t)$, $j=1,\ldots,n_{i};\;i=1,\ldots,k$ which have been introduced in Section~\ref{NumImp}.
		\item[(2)] For each $i=1,\ldots,k$, draw a sample from $\bz_{ij}(t),j=1,\ldots,n_{i}$  with replacement with size $n_i$, and denote these $k$ bootstrap samples as $\tilde{\bz}_{ij}(t),j=1,\ldots,n_{i};\;i=1,\ldots,k$.
		\item[(3)] Compute the bootstrapped test statistic $\tilde{T}_{n}$ based on the bootstrap samples $\tilde{\bz}_{ij}(t),j=1,\ldots,n_{i};\;i=1,\ldots,k$ given in Step (2).
		\item[(4)] Repeat Steps (2)--(3) $B$ times so that $B$ bootstrapped test statistics $\tilde{T}_{n}^{(1)},\ldots,\tilde{T}_{n}^{(B)}$ are obtained.
		\item[(5)] The $p$-value of the bootstrap test is calculated as $B^{-1}\sum_{b=1}^BI\{\tilde{T}_{n}^{(b)}>T_n\}$ where $I\{A\}$ denotes the indicator function of $A$.
	\end{enumerate}
Note that the null hypothesis (\ref{H0G5.equ}) holds whatever $\delta$ is. We only compare the performance of $T_{\NEW}$ and $T_{\BT}$ in terms of size control. Throughout this paper, we set $B=1000$.
\end{enumerate}
\begin{table}[!h]
	\setlength{\tabcolsep}{1pt} 
	\caption{Simulation 1. Empirical sizes (in $\%$) of W, LH, P, R, $T_{\ZZC}^{\N}$, $T_{\ZZC}^{\B}$, and $T_{\NEW}$  of H1.}  \label{size1.tab}	
	\centering
	\begin{tabular*}{\textwidth}{@{\extracolsep{\fill}}cccccccccc}
		\toprule
		$\rho$ & $\varepsilon_{ijr}$ & $\bn$     & W     & LH    & P     & R     &$T_{\ZZC}^{\N}$ & $T_{\ZZC}^{\B}$  & $T_{\NEW}$  \\
		\cmidrule{4-10}
		\multirow{12}[0]{*}{$0.1$}  &\multirow{4}[0]{*}{$N(0,1)$} & $\bn_1$    & 4.0   & 4.7   & 3.1   & 6.7   & 16.2  & 25.7  & 5.6 \\
		&  & $\bn_2$    & 4.2   & 4.7   & 3.8   & 8.5   & 11.0  & 14.1  & 5.7 \\
		&& $\bn_3$    & 4.7   & 4.8   & 4.1   & 7.6   & 8.4   & 10.1  & 5.8 \\
		&& $\bn_4$    & 3.8   & 3.9   & 3.3   & 7.3   & 5.6   & 6.3   & 5.0 \\
		
		& \multirow{4}[0]{*}{$t_4/\sqrt{2}$} & $\bn_1$    & 3.2   & 4.0   & 2.9   & 7.3   & 18.3  & 28.4  & 4.3 \\
		& & $\bn_2$    & 3.9   & 4.4   & 3.3   & 7.6   & 10.1  & 15.8  & 5.3 \\
		&& $\bn_3$    & 4.2   & 4.2   & 4.1   & 8.1   & 8.0   & 9.7   & 5.0 \\
		&& $\bn_4$    & 3.4   & 3.5   & 3.1   & 7.0   & 5.9   & 7.4   & 4.7 \\
		
		&\multirow{4}[0]{*}{\large $\frac{\chi_4^2-4}{2\sqrt{2}}$} & $\bn_1$    & 4.6   & 5.3   & 3.9   & 7.5   & 18.7  & 28.2  & 6.1 \\
		&& $\bn_2$    & 5.5   & 5.9   & 4.7   & 9.7   & 11.7  & 17.4  & 6.7 \\
		&  & $\bn_3$    & 5.2   & 5.1   & 4.6   & 8.6   & 8.5   & 10.4  & 6.9 \\
		& & $\bn_4$    & 3.6   & 3.7   & 3.5   & 7.4   & 6.4   & 7.1   & 4.9 \\
		\cmidrule{4-10}
		\multicolumn{3}{c}{ARE} & 18.5  & 14.0  & 26.0  & 55.5  & 114.7 & 201.0 & 13.7\\
		
		\midrule
		$\rho$  &$\varepsilon_{ijr}$ & $\bn$     & W     & LH    & P     & R      &$T_{\ZZC}^{\N}$ & $T_{\ZZC}^{\B}$  & $T_{\NEW}$   \\
		\cmidrule{4-10}
		\multirow{12}[0]{*}{$0.5$} &\multirow{4}[0]{*}{$N(0,1)$} & $\bn_1$   & 1.9   & 2.3   & 1.4   & 5.3   & 18.5  & 39.8  & 6.6 \\
		&  & $\bn_2$    & 2.9   & 3.6   & 2.2   & 7.1   & 9.7   & 19.1  & 6.0 \\
		& & $\bn_3$    & 3.9   & 4.0   & 3.4   & 6.9   & 8.5   & 12.1  & 6.9 \\
		& & $\bn_4$   & 2.6   & 2.9   & 2.5   & 5.2   & 5.7   & 7.3   & 4.8 \\
		
		&\multirow{4}[0]{*}{$t_4/\sqrt{2}$} & $\bn_1$     & 2.0   & 2.4   & 1.5   & 5.2   & 18.1  & 41.7  & 4.6 \\
		& & $\bn_2$     & 2.6   & 3.0   & 2.3   & 7.1   & 8.7   & 20.6  & 4.9 \\
		&& $\bn_3$      & 3.1   & 3.5   & 2.9   & 7.9   & 6.8   & 11.1  & 5.4 \\
		&& $\bn_4$     & 3.1   & 3.2   & 2.8   & 6.6   & 4.9   & 6.9   & 4.3 \\
		&\multirow{4}[0]{*}{\large $\frac{\chi_4^2-4}{2\sqrt{2}}$} & $\bn_1$    & 2.7   & 3.4   & 2.3   & 5.2   & 19.5  & 42.1  & 5.6 \\
		&& $\bn_2$    & 4.5   & 4.8   & 4.5   & 7.2   & 10.3  & 21.3  & 7.0 \\
		&& $\bn_3$    & 4.0   & 4.2   & 3.5   & 8.4   & 8.2   & 12.4  & 6.4 \\
		&& $\bn_4$    & 2.7   & 2.7   & 2.7   & 5.8   & 6.5   & 9.2   & 5.3 \\
		\cmidrule{4-10}
		\multicolumn{3}{c}{ARE} & 40.0  & 33.3  & 46.7  & 29.8  & 109.3 & 306.0 & 17.7 \\
		\midrule
		$\rho$ &$\varepsilon_{ijr}$ & $\bn$     & W     & LH    & P     & R      &$T_{\ZZC}^{\N}$ & $T_{\ZZC}^{\B}$  & $T_{\NEW}$  \\
		\cmidrule{4-10}
		\multirow{12}[0]{*}{$0.9$} &\multirow{4}[0]{*}{$N(0,1)$} & $\bn_1$     & 0.8   & 0.9   & 0.8   & 4.3   & 17.1  & 56.7  & 4.1 \\
		&& $\bn_2$    & 1.4   & 1.4   & 1.4   & 5.4   & 6.9   & 26.4  & 4.3 \\
		&& $\bn_3$    & 2.1   & 2.1   & 2.1   & 5.1   & 6.0   & 13.8  & 5.1 \\
		&& $\bn_4$    & 1.9   & 1.9   & 1.7   & 5.4   & 5.3   & 8.5   & 5.3 \\
		&\multirow{4}[0]{*}{$t_4/\sqrt{2}$} & $\bn_1$    & 0.5   & 0.6   & 0.4   & 4.0   & 15.6  & 58.5  & 3.5 \\
		&& $\bn_2$       & 1.3   & 1.3   & 1.1   & 4.3   & 6.4   & 27.9  & 3.4 \\
		&& $\bn_3$     & 2.6   & 2.6   & 2.5   & 5.5   & 4.9   & 14.4  & 4.0 \\
		&& $\bn_4$       & 2.0   & 2.0   & 2.0   & 5.5   & 4.7   & 9.8   & 4.5 \\
		&\multirow{4}[0]{*}{\large $\frac{\chi_4^2-4}{2\sqrt{2}}$} & $\bn_1$     & 1.4   & 1.4   & 1.1   & 4.9   & 15.7  & 57.5  & 3.8 \\
		&& $\bn_2$     & 2.4   & 2.5   & 2.3   & 6.5   & 9.1   & 28.3  & 6.7 \\
		&& $\bn_3$      & 2.7   & 2.7   & 2.6   & 5.9   & 7.1   & 16.5  & 6.4 \\
		&& $\bn_4$      & 1.9   & 1.9   & 1.9   & 6.1   & 5.7   & 10.2  & 5.4 \\
		\cmidrule{4-10}
		\multicolumn{3}{c}{ARE} &  65.0	&64.5	&66.8	&13.2	&75.5	&447.5	&18.8\\
		\bottomrule
	\end{tabular*}
\end{table}

Table~\ref{size1.tab} presents the empirical sizes of W, LH, P, R, $T_{\ZZC}^{\N}$, $T_{\ZZC}^{\B}$, and $T_{\NEW}$  of H1, with the last row displaying their  average relative error (ARE) values associated with the three values of $\rho$. Here we adopt the value of ARE by \cite{zhang2011tech} to measure the overall performance of a test in maintaining the nominal size.   It can be calculated as $\ARE=100J^{-1}\sum_{j=1}^J|\hat{\alpha}_j-\alpha|/\alpha$, where $\hat{\alpha}_j,j=1,\ldots,J$ denote the empirical sizes under $J$ simulation settings. A smaller ARE value of a test indicates a better performance of that test in terms of size control. Overall, based on the ARE values, $T_{\NEW}$ performs  reasonably well and outperforms W, P, $T_{\ZZC}^{\N}$ and $T_{\ZZC}^{\B}$, regardless of whether the simulated functional data are highly correlated ($\rho=0.1$), moderately correlated ($\rho=0.5$), or less correlated ($\rho=0.9$). Its ARE values are 13.7, 17.7, and 18.8, respectively, all remaining below 20, consistent with the benchmark established by \cite{zhang2012approximate}, while the ARE values for these four competitors are significantly larger.   $T_{\NEW}$ is comparable to LH when $\rho=0.1$ and slightly worse than R when $\rho=0.9$. In particular, W, LH, and P are very conservative with small samples and especially when $\rho$ is large. This is expected as these tests were developed under the assumption of homoscedasticity. When $\rho$ is large, heteroscedasticity among the three samples increases, which causes these tests to perform poorly.  In addition, both $T_{\ZZC}^{\N}$ and $T_{\ZZC}^{\B}$ are very liberal, particularly with small sample sizes. This is not surprising, as we are working with smaller sample sizes compared to the simulation studies  in \cite{zhu2023ksampFMD}. With smaller sample sizes, the discrepancy between using $\bOmega_n^{-1}(s,t)$ and $\hbOmega_n^{-1}(s,t)$ becomes more pronounced, leading to a notable bias. Consequently, the performance of $T_{\ZZC}^{\B}$ is less than optimal. It is seen from Table~\ref{size1.tab} that the performance of $T_{\ZZC}^{\N}$ and  $T_{\ZZC}^{\B}$ improves as the sample size increases. In contrast, by incorporating the adjustment coefficient $c_n$ (\ref{cn.equ}), which considers both the sample sizes and the correlation among the simulated functional data, the performance of $T_{\NEW}$ is enhanced.  Although it is somewhat conservative when the sample size is very small ($\bn=[7,12,12]$) and $\rho=0.9$, it remains effective even with relatively small sample sizes and varying levels of correlation among the simulated functional data.

For the alternative hypothesis, an anonymous reviewer suggested that comparing the powers of tests is inappropriate if the tests are either too conservative or too liberal. Therefore, for the power comparison, we focus only on scenarios with large sample size to ensure that the tests under consideration are neither too conservative nor too liberal, i.e., $\bn=[50,70,70]$. Since the functional samples  become more noisy if the value of $\rho$ becomes larger, the empirical power of a test becomes smaller when $\rho$  is larger for the same value of $\delta$.  Accordingly, we set $\delta=0.15,0.2,0.25,0.3$ when $\rho=0.1$; $\delta=0.2,0.4,0.6,0.8$ when $\rho=0.5$; and $\delta=0.8,1.0,1.2,1.4$ when $\rho=0.9$. 

\begin{table}[!h]
  \setlength{\tabcolsep}{1pt} 
  \caption{Simulation 1. Empirical powers (in $\%$) of W, LH, P, R, $T_{\ZZC}^{\N}$, $T_{\ZZC}^{\B}$, and $T_{\NEW}$  of H1 when $\bn=[50,70,70]$.}\label{power1.tab}	
    \centering
   \begin{tabular*}{\textwidth}{@{\extracolsep{\fill}}cclcccccccc}
   \toprule
     $\rho$ & $\varepsilon_{ijr}$ &$\delta$    & W     & LH    & P     & R    &$T_{\ZZC}^{\N}$ & $T_{\ZZC}^{\B}$  & $T_{\NEW}$  \\
\midrule
\multirow{12}[0]{*}{0.1} & \multirow{4}[0]{*}{$N(0,1)$} & 0.15  & 13.5  & 13.5  & 13.3  & 16.0  & 18.5  & 19.9  & 16.7 \\
          &       & 0.2   & 26.0  & 26.1  & 25.9  & 28.5  & 36.1  & 38.5  & 31.8 \\
          &       & 0.25  & 48.9  & 48.7  & 48.6  & 49.0  & 60.3  & 62.4  & 56.5 \\
          &       & 0.3   & 74.1  & 74.3  & 74.0  & 73.9  & 84.3  & 86.1  & 80.4 \\
          & \multirow{4}[0]{*}{$t_4/\sqrt{2}$} & 0.15  & 17.0  & 17.2  & 16.8  & 19.2  & 24.1  & 26.5  & 20.4 \\
          &       & 0.2   & 32.7  & 33.0  & 33.4  & 32.6  & 41.0  & 45.3  & 37.8 \\
          &       & 0.25  & 55.3  & 55.2  & 55.6  & 55.7  & 66.2  & 69.1  & 61.1 \\
          &       & 0.3   & 79.1  & 79.0  & 78.9  & 77.8  & 85.0  & 87.4  & 82.2 \\
          & \multirow{4}[0]{*}{\large $\frac{\chi_4^2-4}{2\sqrt{2}}$} & 0.15  & 12.7  & 12.7  & 13.1  & 14.9  & 18.7  & 21.5  & 16.5 \\
          &       & 0.2  & 26.4  & 26.5  & 26.3  & 24.7  & 34.0  & 37.8  & 29.9 \\
          &       & 0.25  & 49.3  & 49.1  & 49.2  & 46.4  & 60.2  & 64.2  & 55.4 \\
          &       & 0.3 & 78.3  & 78.0  & 78.1  & 74.4  & 84.7  & 87.6  & 82.5 \\
          \midrule
    \multirow{12}[0]{*}{0.5} & \multirow{4}[0]{*}{$N(0,1)$} & 0.2   & 4.4   & 4.6   & 4.4   & 8.3   & 9.2   & 12.1  & 8.2 \\
          &       & 0.4   & 16.4  & 16.4  & 16.2  & 22.8  & 26.6  & 30.6  & 23.7 \\
          &       & 0.6   & 46.3  & 47.0  & 45.9  & 57.4  & 64.5  & 68.5  & 60.8 \\
          &       & 0.8   & 86.3  & 86.8  & 85.8  & 91.5  & 93.5  & 95.4  & 92.5 \\
          & \multirow{4}[0]{*}{$t_4/\sqrt{2}$} & 0.2   & 4.9   & 4.9   & 4.9   & 10.0  & 10.1  & 13.2  & 8.9 \\
          &       & 0.4   & 19.1  & 19.3  & 18.3  & 23.3  & 29.6  & 35.7  & 27.0 \\
          &       & 0.6   & 51.9  & 52.5  & 51.1  & 62.4  & 66.3  & 73.3  & 63.5 \\
          &       & 0.8   & 88.6  & 88.6  & 88.2  & 92.0  & 94.0  & 96.3  & 93.1 \\
          & \multirow{4}[0]{*}{\large $\frac{\chi_4^2-4}{2\sqrt{2}}$} & 0.2   & 4.7   & 5.0   & 4.6   & 8.6   & 9.4   & 12.7  & 8.3 \\
          &       & 0.4   & 15.9  & 16.4  & 15.9  & 21.1  & 25.8  & 31.7  & 23.7 \\
          &       & 0.6   & 49.4  & 49.7  & 48.8  & 56.8  & 64.6  & 71.8  & 61.8 \\
          &       & 0.8   & 89.4  & 89.5  & 89.4  & 94.0  & 94.8  & 95.9  & 94.3 \\
          \midrule
    \multirow{12}[0]{*}{0.9} & \multirow{4}[0]{*}{$N(0,1)$} & 0.8   & 18.0  & 18.3  & 17.9  & 36.7  & 33.5  & 45.2  & 32.2 \\
          &       & 1.0     & 39.5  & 40.0  & 38.8  & 63.2  & 56.1  & 66.4  & 54.9 \\
          &       & 1.2   & 63.3  & 63.9  & 62.2  & 87.0  & 78.3  & 86.1  & 77.8 \\
          &       & 1.4   & 85.1  & 85.7  & 84.6  & 97.6  & 92.9  & 96.8  & 92.4 \\
          & \multirow{4}[0]{*}{$t_4/\sqrt{2}$} & 0.8   & 21.5  & 21.9  & 21.4  & 40.8  & 35.2  & 48.3  & 34.2 \\
          &       & 1.0     & 41.5  & 41.9  & 41.1  & 65.3  & 57.7  & 70.6  & 56.6 \\
          &       & 1.2   & 66.4  & 66.9  & 65.9  & 87.7  & 79.1  & 88.5  & 78.1 \\
          &       & 1.4   & 88.4  & 88.6  & 88.1  & 97.0  & 93.3  & 96.9  & 92.9 \\
          & \multirow{4}[0]{*}{\large $\frac{\chi_4^2-4}{2\sqrt{2}}$} & 0.8   & 19.2  & 19.2  & 18.9  & 35.9  & 32.5  & 46.6  & 31.0 \\
          &       & 1.0     & 39.8  & 40.3  & 39.2  & 63.5  & 58.0  & 70.0  & 56.9 \\
          &       & 1.2   & 66.7  & 67.0  & 65.9  & 88.4  & 81.6  & 89.6  & 80.1 \\
          &       & 1.4   & 88.4  & 89.1  & 87.6  & 98.0  & 94.8  & 97.2  & 94.6 \\
 \bottomrule
    \end{tabular*}%
\end{table}%

The empirical powers (in $\%$) of W, LH, P, R, $T_{\ZZC}^{\N}$, $T_{\ZZC}^{\B}$, and $T_{\NEW}$   of H1 for $\bn=[50,70,70]$ are presented in Table~\ref{power1.tab}.  As expected, increasing the value of $\delta$ leads to an increase in the empirical power of a test. Moreover, larger sample sizes can yield higher empirical powers. Table~\ref{power1.tab} shows that the empirical powers of W, LH, and P remain consistently lower than those of the other four tests,  due to their conservative behavior observed in Table~\ref{size1.tab} even with a large sample size. Additionally,  the empirical powers of $T_{\ZZC}^{\B}$  are consistently the highest across all settings as it tends to exhibit a somewhat liberal behavior even with a large sample size.  For R, $T_{\ZZC}^{\N}$, and $T_{\NEW}$, comparable empirical sizes lead to comparable empirical powers.

\begin{table}[!h]  
   \setlength{\tabcolsep}{1pt} 
	\caption{Simulation 1. Empirical sizes and powers (in $\%$) of $T_{\QCZ}$, $T_{\QCZ}^{\max}$, and $T_{\NEW}$  of H2 when $\bn=[50,70,70]$.}  \label{size2.tab}	
			\begin{tabular*}{\textwidth}{@{\extracolsep{\fill}}cclccclccclccc}
				\toprule%
				&       & \multicolumn{4}{c}{$\rho=0.1$} & \multicolumn{4}{c}{$\rho=0.5$} & \multicolumn{4}{c}{$\rho=0.9$} \\          
				\cmidrule{3-6} \cmidrule{7-10} \cmidrule{11-14}
			$\bG$	&$\varepsilon_{ijr}$    & $\delta$     &$T_{\QCZ}$  & $T_{\QCZ}^{\max}$  & $T_{\NEW}$   & $\delta$     &$T_{\QCZ}$  & $T_{\QCZ}^{\max}$  & $T_{\NEW}$    & $\delta$    &$T_{\QCZ}$  & $T_{\QCZ}^{\max}$  & $T_{\NEW}$ 
				\\    
   \midrule
    \multirow{15}[0]{*}{$\bG_2$} & \multirow{5}[0]{*}{$N(0,1)$}  & 0  & 3.5   & 1.7   & 4.4   & 0  & 4.9   & 3.9   & 7.6   & 0   & 2.3   & 2.8   & 6.5 \\
          &       & 0.15  & 13.8  & 31.6  & 16.1  & 0.2   & 5.6   & 4.2   & 8.9   & 0.8   & 14.3  & 13.1  & 26.6 \\
          &       & 0.2  & 21.1  & 62.9  & 25.6  & 0.4   & 14.9  & 19.9  & 20.8  & 1.0   & 25.1  & 20.6  & 39.8 \\
          &       & 0.25  & 34.8  & 85.6  & 38.9  & 0.6   & 35.3  & 58.3  & 45.9  & 1.2   & 38.3  & 37.1  & 55.6 \\
          &       & 0.3  & 53.8  & 97.8  & 59.4  & 0.8   & 66.4  & 90.8  & 73.9  & 1.4   & 61.3  & 57.5  & 75.2 \\
          & \multirow{5}[0]{*}{$t_4/\sqrt{2}$} & 0  & 3.9   & 3.3   & 5.0   & 0   & 3.7   & 3.0   & 5.8   & 0  & 1.6   & 2.5   & 5.3 \\
          &       & 0.15  & 14.0  & 34.4  & 16.3  & 0.2   & 6.3   & 5.7   & 9.5   & 0.8   & 12.7  & 10.9  & 21.7 \\
          &       & 0.2  & 24.3  & 67.0  & 28.6  & 0.4   & 14.2  & 18.5  & 20.1  & 1.0   & 26.0  & 23.2  & 40.3 \\
          &       & 0.25  & 34.6  & 89.3  & 38.9  & 0.6   & 36.5  & 58.1  & 44.1  & 1.2   & 41.4  & 40.9  & 56.4 \\
          &       & 0.3  & 54.2  & 98.5  & 61.0  & 0.8   & 67.3  & 91.8  & 74.3  & 1.4   & 62.1  & 60.5  & 76.5 \\
          & \multirow{5}[0]{*}{\large $\frac{\chi_4^2-4}{2\sqrt{2}}$} & 0  & 4.5   & 3.0   & 5.0   & 0   & 3.5   & 3.3   & 6.0   & 0   & 1.9   & 2.6   & 5.3 \\
          &       & 0.15  & 10.8  & 28.3  & 14.5  & 0.2   & 5.0   & 4.3   & 8.3   & 0.8   & 14.9  & 13.2  & 26.1 \\
          &       & 0.2  & 19.0  & 61.3  & 24.6  & 0.4   & 12.7  & 16.2  & 18.5  & 1.0   & 25.2  & 21.6  & 39.3 \\
          &       & 0.25  & 29.9  & 88.8  & 35.2  & 0.6   & 34.6  & 55.8  & 43.8  & 1.2   & 39.9  & 39.3  & 55.3 \\
          &       & 0.3  & 52.1  & 98.9  & 58.3  & 0.8   & 68.4  & 90.4  & 75.2  & 1.4   & 62.4  & 57.0  & 75.4 \\
          \midrule
    \multirow{15}[0]{*}{$\bG_3$} & \multirow{5}[0]{*}{$N(0,1)$} & 0  & 1.7   & 1.7   & 5.4   & 0   & 1.2   & 1.0   & 5.3   & 0   & 0.4   & 0.9   & 5.9 \\
          &       & 0.15  & 18.6  & 66.6  & 29.7  & 0.2   & 5.4   & 5.7   & 14.0  & 0.8   & 22.7  & 28.5  & 54.2 \\
          &       & 0.2  & 38.7  & 94.7  & 54.4  & 0.4   & 22.1  & 41.4  & 41.2  & 1.0   & 48.0  & 51.8  & 78.1 \\
          &       & 0.25  & 68.3  & 99.9  & 80.4  & 0.6   & 65.4  & 92.7  & 82.2  & 1.2   & 77.3  & 79.1  & 95.3 \\
          &       & 0.3  & 86.8  & 100.0 & 94.1  & 0.8   & 95.7  & 99.9  & 99.1  & 1.4   & 93.2  & 96.0  & 99.3 \\
          & \multirow{5}[0]{*}{$t_4/\sqrt{2}$} & 0  & 2.6   & 2.2   & 6.2   & 0   & 1.3   & 1.7   & 5.5   & 0   & 0.5   & 1.1   & 5.1 \\
          &       & 0.15  & 18.9  & 68.6  & 30.6  & 0.2   & 4.8   & 5.6   & 12.0  & 0.8   & 22.0  & 28.4  & 55.2 \\
          &       & 0.2  & 43.8  & 95.2  & 58.3  & 0.4   & 23.8  & 47.0  & 43.6  & 1.0   & 48.7  & 58.2  & 81.2 \\
          &       & 0.25  & 68.4  & 99.9  & 81.8  & 0.6   & 70.0  & 95.3  & 87.4  & 1.2   & 78.8  & 83.5  & 94.7 \\
          &       & 0.3  & 89.4  & 100.0 & 94.8  & 0.8   & 95.8  & 100.0 & 99.1  & 1.4   & 94.3  & 95.9  & 99.0 \\
          & \multirow{5}[0]{*}{\large $\frac{\chi_4^2-4}{2\sqrt{2}}$} & 0 & 3.8   & 2.8   & 6.1   & 0   & 2.0   & 2.0   & 6.5   & 0   & 0.3   & 1.3   & 4.8 \\
          &       & 0.15  & 17.1  & 63.5  & 28.4  & 0.2   & 3.4   & 3.9   & 11.5  & 0.8   & 22.0  & 25.0  & 56.6 \\
          &       & 0.2  & 34.4  & 95.3  & 51.7  & 0.4   & 20.4  & 40.3  & 38.1  & 1.0   & 50.9  & 55.2  & 81.6 \\
          &       & 0.25  & 67.4  & 99.8  & 82.2  & 0.6   & 66.9  & 92.9  & 85.7  & 1.2   & 78.4  & 82.1  & 96.2 \\
          &       & 0.3  & 90.8  & 100.0 & 96.4  & 0.8   & 95.4  & 100.0 & 98.9  & 1.4   & 93.4  & 96.3  & 99.8 \\
          \midrule
    \multirow{15}[0]{*}{$\bG_4$} & \multirow{5}[0]{*}{$N(0,1)$} & 0  & 7.8   & 5.7   & 5.6   & 0   & 7.2   & 5.0   & 5.7   & 0   & 5.8   & 5.3   & 5.3 \\
          &       & 0.15  & 11.3  & 18.7  & 9.2   & 0.2   & 7.9   & 6.5   & 6.4   & 0.8   & 14.6  & 10.0  & 12.6 \\
          &       & 0.2  & 16.0  & 32.6  & 13.6  & 0.4   & 14.1  & 13.1  & 12.0  & 1.0   & 22.5  & 17.5  & 20.9 \\
          &       & 0.25  & 23.1  & 56.9  & 20.1  & 0.6   & 25.6  & 31.6  & 22.9  & 1.2   & 31.3  & 23.6  & 29.3 \\
          &       & 0.3  & 30.7  & 76.7  & 25.7  & 0.8   & 42.1  & 60.1  & 38.3  & 1.4   & 41.4  & 33.1  & 38.3 \\
          & \multirow{5}[0]{*}{$t_4/\sqrt{2}$} & 0  & 7.4   & 6.7   & 6.2   & 0   & 6.1   & 5.9   & 5.0   & 0   & 5.9   & 6.2   & 5.1 \\
          &       & 0.15  & 13.7  & 21.4  & 11.5  & 0.2   & 8.4   & 7.6   & 6.7   & 0.8   & 14.3  & 12.5  & 12.9 \\
          &       & 0.2  & 16.7  & 37.4  & 14.6  & 0.4   & 13.1  & 13.2  & 11.5  & 1.0   & 20.9  & 16.8  & 18.1 \\
          &       & 0.25  & 24.3  & 58.3  & 21.0  & 0.6   & 25.3  & 32.8  & 22.8  & 1.2   & 30.6  & 25.8  & 28.0 \\
          &       & 0.3  & 31.8  & 80.4  & 28.3  & 0.8   & 39.9  & 60.8  & 36.3  & 1.4   & 39.7  & 34.7  & 36.9 \\
          & \multirow{5}[0]{*}{\large $\frac{\chi_4^2-4}{2\sqrt{2}}$} & 0  & 7.7   & 6.7   & 6.4   & 0   & 6.0   & 6.2   & 5.3   & 0   & 6.8   & 6.7   & 5.8 \\
          &       & 0.15  & 9.7   & 14.7  & 7.5   & 0.2   & 7.5   & 6.0   & 5.9   & 0.8   & 16.0  & 12.4  & 14.7 \\
          &       & 0.2  & 13.6  & 28.4  & 11.6  & 0.4   & 11.5  & 10.3  & 9.8   & 1.0   & 21.5  & 15.6  & 19.1 \\
          &       & 0.25  & 22.8  & 55.1  & 19.1  & 0.6   & 23.3  & 28.5  & 20.0  & 1.2   & 27.9  & 24.5  & 26.3 \\
          &       & 0.3  & 30.0  & 78.4  & 26.1  & 0.8   & 40.4  & 56.9  & 36.7  & 1.4   & 39.4  & 33.3  & 35.7 \\
   \bottomrule
		\end{tabular*}
\end{table}

When the null hypothesis (\ref{GLHTH0.sec2}) of the GLHT problem is rejected, it is often of interest to further conduct  some contrast tests. Now we are targeting to examine the finite-sample performance of $T_{\NEW}$ of H2, that is, the two-sample problem for MFD. To ensure a fair comparison with the simulation results in \cite{Qiu2021}, which used a sample size of $\bn = [60, 90]$, we adopt a similar sample size and present the empirical sizes and powers  (in \%) of $T_{\QCZ}$, $T_{\QCZ}^{\max}$, and $T_{\NEW}$ for $\bn=[50,70,70]$ in Table~\ref{size2.tab}. Several observations can be made from this table. Firstly, when the coefficient matrix $\bG$ is specified as $\bG_2$, that is, we are comparing the two mean functions $\bfeta_1(t)$  and $\bfeta_2(t)$, $T_{\NEW}$ generally performs well regardless of the level of correlation among the simulated functional data ($\rho=0.1$, $0.5$, or $0.9$), with empirical sizes ranging from 4.4\% to 7.6\%.  $T_{\QCZ}$ tends to be slightly conservative, with empirical sizes ranging from 1.6\% to 4.9\%. $T_{\QCZ}^{\max}$  is even more conservative than $T_{\NEW}$ and $T_{\QCZ}$, particularly when $\rho=0.9$. Secondly, when $\bG=\bG_3$, that is, we are comparing the two mean functions, $\bfeta_1(t)$ and $\bfeta_3(t)$, both $T_{\QCZ}$ and $T_{\QCZ}^{\max}$ are more conservative than $T_{\NEW}$. This is not surprising since  $T_{\QCZ}$ and $T_{\QCZ}^{\max}$ are proposed under the assumption of a homogeneous covariance function, which is strongly violated when $\rho=0.5$ and $0.9$. However, this disadvantage can be overcome when the sample sizes of two groups are equal. When $\bG=\bG_3$,  that is, we are comparing the two mean functions, $\bfeta_2(t)$ and $\bfeta_3(t)$, since we set $n_2=n_3$, it is evident from Table~\ref{size2.tab}, $T_{\QCZ}$ and $T_{\QCZ}^{\max}$ are  no more as conservative as in the previous two cases. Thirdly, in terms of power, it is seen that when the functional data are highly correlated ($\rho=0.1$) and moderately correlated ($\rho=0.5$), $T_{\QCZ}^{\max}$ outperforms the other two tests, and $T_{\NEW}$ and $T_{\QCZ}$ show comparable performance. $T_{\NEW}$ outperforms $T_{\QCZ}$ and $T_{\QCZ}^{\max}$ when the simulated functional samples are less correlated $(\rho=0.9$) and the two sample sizes are different. These are also consistent with the results of \cite{smaga2019linear}. Nevertheless, the application of $T_{\NEW}$ is boarder compared to $T_{\QCZ}$ and $T_{\QCZ}^{\max}$ as the latter two are designed specifically for two-sample problems.

Last, we examine the finite-sample performance for a specific linear hypothesis, H3, via comparing it with the bootstrap-based test $T_{\BT}$. The empirical sizes of $T_{\BT}$ and $T_{\NEW}$  of H3 are presented in Table~\ref{size3.tab} with the last row displaying their ARE values associated the three cases of $\rho$.  It is seen that $T_{\BT}$ is rather conservative especially when $\rho$ is large. This observation highlights that the bootstrap-based test dose not work well when dealing with the simulated functional data which are less correlated. Moreover, the bootstrap-based tests are commonly recognized as time-consuming methods. In contrast, $T_{\NEW}$ generally performs well, though it exhibits slight liberal tendencies when the sample size is very small ($\bn=[7,12,12]$). However, it consistently outperforms $T_{\BT}$ as sample sizes increase and for less correlated simulated functional data. Notably, $T_{\NEW}$ offers a significant computational advantage, being substantially faster than the bootstrap-based test.

\begin{table}[!h]  
	\caption{Simulation 1. Empirical sizes (in $\%$) of $T_{\BT}$ and $T_{\NEW}$  of H3.}  \label{size3.tab}	
			\begin{tabular*}{\textwidth}{@{\extracolsep{\fill}}cccccccc}
				\toprule
				&       & \multicolumn{2}{c}{$\rho=0.1$} & \multicolumn{2}{c}{$\rho=0.5$} & \multicolumn{2}{c}{$\rho=0.9$} \\          
			\cmidrule{3-4} \cmidrule{5-6} \cmidrule{7-8}
            	$\varepsilon_{ijr}$ & $\bn$     &$T_{\BT}$    & $T_{\NEW}$    &$T_{\BT}$    & $T_{\NEW}$   &$T_{\BT}$    & $T_{\NEW}$ 	\\    
			  \cmidrule{3-8}
    \multirow{4}[0]{*}{$N(0,1)$} & $\bn_1$  & 1.7   & 5.9   & 1.0   & 7.6   & 0.2   & 6.3 \\
          & $\bn_2$    & 3.1   & 6.5   & 1.9   & 6.9   & 0.8   & 7.1 \\
          & $\bn_3$    & 5.9   & 6.6   & 5.4   & 7.0   & 2.9   & 6.8 \\
          & $\bn_4$    & 5.5   & 6.3   & 5.0   & 6.3   & 3.6   & 6.4 \\
    \multirow{4}[0]{*}{$t_4/\sqrt{2}$} & $\bn_1$  & 1.3   & 5.9   & 0.3   & 5.7   & 0.1   & 4.3 \\
          & $\bn_2$    & 2.1   & 6.3   & 1.2   & 6.4   & 0.6   & 5.4 \\
          & $\bn_3$    & 3.5   & 5.0   & 3.0   & 5.4   & 2.0   & 4.7 \\
          & $\bn_4$    & 3.8   & 4.4   & 3.7   & 5.5   & 3.6   & 5.6 \\
    \multirow{4}[0]{*}{\large $\frac{\chi_4^2-4}{2\sqrt{2}}$} & $\bn_1$    & 1.1   & 7.0   & 0.7   & 8.3   & 0.1   & 4.7 \\
          & $\bn_2$    & 3.4   & 6.7   & 2.4   & 7.6   & 1.0   & 6.2 \\
          & $\bn_3$    & 5.2   & 6.7   & 4.0   & 6.7   & 3.1   & 6.3 \\
          & $\bn_4$    & 5.5   & 6.0   & 4.7   & 5.8   & 3.9   & 5.3 \\
          \cmidrule{3-8}
    \multicolumn{2}{c}{ARE} &36.8	&24.2	&45.8	&32.0	&63.5	&19.5\\
    \bottomrule
    \end{tabular*}%
\end{table}%

In conclusion, we have conducted various hypothesis tests in this simulation by setting various coefficient matrix $\bG$. The results demonstrate that our proposed test, $T_{\NEW}$, generally performs reasonably well  no matter how the functional data correlated and does not require too large sample size to achieve good performance. The broader applicability of $T_{\NEW}$ suggests that it can be employed in a wider range of scenarios or research contexts beyond the specific framework proposed by \cite{Qiu2021,zhu2023ksampFMD}. This flexibility allows researchers to adapt and use $T_{\NEW}$ based on their specific needs and study designs.

\subsection{Simulation 2}
In this simulation study,  we demonstrate the good performance of $T_{\NEW}$ in more comprehensive and practical cases. As mentioned in Section~\ref{NumImp}, in practice, the functional data are usually observed at a grid of design time points and these time points may vary across different observations. In such a situation, it is necessary to initially reconstruct the functional data using a smoothing technique and subsequently discretize each reconstructed function at a shared set of design time points. The proposed global test then can be applied to the reconstructed data.  We consider $k=4$ groups of multivariate functional samples with three vectors $\bn=[n_1,n_2,n_3,n_4]$ of sample sizes $\bn_1=[7,10,12,15]$, $\bn_2=[15,20,25,30]$, and $\bn_3=[20,25,30,40]$. The four functional observations are generated from the model which has been similarly considered by  \cite{gorecki2017multivariate} (M1 in Section 3.1):
\begin{equation}\label{modelS2.equ}
\by_{ij}(t)=\bfeta_i(t)+\bA_i(B_1,B_2)^\top,i=1,\ldots,4,
\end{equation}
where  $B_1$ and $B_2$ are independent standard Brownian motions with dispersion parameter $0.2^2$. The vector of mean functions are set as $\eta_{i1}(t)=[\sin(2\pi t^2)]^5$ and $\eta_{i2}(t)=t^{1/5}(1-t)^{6-1/5}-5$ for $i=1,\ldots,4$. \mrk{The matrix $\bA_i$ captures the covariance structure among the components of $\by_{ij}(t)$. If $\bA_i$	is fixed across all groups, we have the homoscedastic case. In contrast,  if $\bA_i$ varies across groups, we have the heteroscedastic case.} 
As recommended by an anonymous reviewer, we consider both homoscedastic and heteroscedastic cases in this simulation study. For the homoscedastic case, we set $\bA_i=0.7\bI_2+0.3\bm{1}_2\bm{1}_2^\top, i=1,\ldots,4$; for the heteroscedastic case, we set $\bA_1=0.7\bI_2+0.3\bm{1}_2\bm{1}_2^\top$, $\bA_2=0.5\bI_2+0.5\bm{1}_2\bm{1}_2^\top$, $\bA_3=0.3\bI_2+0.7\bm{1}_2\bm{1}_2^\top$, and $\bA_4=0.1\bI_2+0.9\bm{1}_2\bm{1}_2^\top$. We aim to evaluate the behavior of our proposed test $T_{\NEW}$ against the four permutation tests proposed by  \cite{gorecki2017multivariate}, namely, W, P, LH, and R,  as well as $T_{\ZZC}^{\N}$ and $T_{\ZZC}^{\B}$ proposed by \cite{zhu2023ksampFMD}, under two scenarios. 
The two scenarios can be described as follows:
\begin{enumerate}[S1.]
	\item[S1.] Model (\ref{modelS2.equ}) with measurement error. That is, $\by_{ij}(t)=\bfeta_i(t)+\bA_i(B_1,B_2)^\top+\bm{e}_{ij},i=1,\ldots,4$, where $e_{ij\ell},\ell=1,2$ are i.i.d.  normally random distributed random variables with mean zero and variance $\sigma^2$. We consider three cases of $\sigma=0.1,0.5,0.9$.
	\item[S2.] First, we generate the MFD from model (\ref{modelS2.equ}). Then, for each observation, we randomly select $aM$ points from its value with
	$a =0.1,  0.5, 0.9$ to roughly generate sparse ($a = 0.1$), semi-dense ($a = 0.5$), and dense ($a = 0.9$) MFD.
\end{enumerate}
Note that for both S1 and S2, we first reconstruct individual functions by using smoothing splines, and then apply  W, LH, P, R, $T_{\ZZC}^{\N}$, $T_{\ZZC}^{\B}$, and $T_{\NEW}$ to the functional samples evaluated at the design time points specified in Simulation 1. Since all the competitors were proposed for the one-way FMANOVA problem (\ref{MANOVA.sec2}), we set $\bG=(\bI_3,-\bm{1}_3)$.

\begin{table}[!h]  
	\caption{\em Simulation 2. Empirical sizes (in $\%$) of  W, LH, P, R, $T_{\ZZC}^{\N}$, $T_{\ZZC}^{\B}$, and $T_{\NEW}$ under S1. }  \label{size4.tab}	
		\begin{tabular*}{\textwidth}{@{\extracolsep{\fill}}cccccccccc@{\extracolsep{\fill}}}
			\toprule
            \multicolumn{9}{l}{Homoscedastic cases}\\
          \midrule
			$\sigma$	& $\bn$        & W     & LH    & P     & R     & $T_{\ZZC}^{\N}$ & $T_{\ZZC}^{\B}$  & $T_{\NEW}$ \\
			\cmidrule{3-9}
    \multirow{3}[0]{*}{0.1} & $\bn_1$    & 4.3   & 4.3   & 4.4   & 3.9   & 5.0   & 8.6   & 4.7 \\
          & $\bn_2$    & 4.9   & 5.0   & 4.7   & 4.9   & 4.8   & 6.2   & 5.0 \\
          & $\bn_3$    & 3.7   & 3.7   & 4.1   & 3.8   & 4.3   & 5.1   & 4.2 \\
    \multirow{3}[0]{*}{0.5} & $\bn_1$    & 4.3   & 4.4   & 4.2   & 4.0   & 5.0   & 7.7   & 4.4 \\
          & $\bn_2$    & 4.5   & 4.4   & 4.3   & 4.5   & 4.7   & 6.0   & 4.6 \\
          & $\bn_3$    & 4.2   & 4.0   & 4.2   & 4.2   & 4.0   & 5.1   & 4.2 \\
    \multirow{3}[0]{*}{0.9} & $\bn_1$    & 4.0   & 4.1   & 4.2   & 3.9   & 4.9   & 8.5   & 4.1 \\
          & $\bn_2$    & 4.0   & 4.4   & 3.9   & 4.7   & 4.4   & 5.9   & 4.2 \\
          & $\bn_3$    & 4.5   & 4.4   & 4.6   & 4.1   & 4.5   & 5.4   & 4.4 \\
         \cmidrule{3-9}
    \multicolumn{2}{c}{ARE} & 14.7  & 14.0  & 14.2  & 15.6  & 7.6   & 30.0  & 11.6 \\
 \midrule
\multicolumn{9}{l}{Heterscedastic cases}\\
\midrule
  $\sigma$	& $\bn$        & W     & LH    & P     & R     & $T_{\ZZC}^{\N}$ & $T_{\ZZC}^{\B}$  & $T_{\NEW}$ \\
			\cmidrule{3-9}
    \multirow{3}[0]{*}{0.1} & $\bn_1$    & 10.0  & 10.3  & 10.4  & 11.1  & 7.4   & 10.8  & 6.0 \\
          & $\bn_2$    & 11.4  & 11.2  & 11.5  & 11.2  & 5.3   & 6.9   & 4.5 \\
          & $\bn_3$    & 9.8   & 9.7   & 9.7   & 9.1   & 5.6   & 6.3   & 4.9 \\
    \multirow{3}[0]{*}{0.5} & $\bn_1$    & 8.9   & 8.6   & 8.9   & 8.8   & 6.3   & 10.7  & 5.6 \\
          & $\bn_2$    & 9.0   & 9.0   & 8.9   & 9.4   & 5.3   & 7.3   & 5.2 \\
          & $\bn_3$    & 8.5   & 8.4   & 8.6   & 7.3   & 5.1   & 6.1   & 4.9 \\
    \multirow{3}[0]{*}{0.9} & $\bn_1$    & 7.1   & 7.0   & 6.9   & 7.2   & 5.6   & 10.0  & 5.3 \\
          & $\bn_2$    & 6.7   & 6.6   & 6.7   & 7.3   & 4.8   & 7.0   & 4.7 \\
          & $\bn_3$    & 6.7   & 6.7   & 6.8   & 6.1   & 4.9   & 6.1   & 4.6 \\
           \cmidrule{3-9}
    \multicolumn{2}{c}{ARE} & 73.6  & 72.2  & 74.2  & 72.2  & 13.1  & 58.2  & 7.8 \\
 	\bottomrule
		\end{tabular*}
\end{table}

The empirical sizes (in $\%$) of  W, LH, P, R,  $T_{\ZZC}^{\N}$, $T_{\ZZC}^{\B}$, and $T_{\NEW}$ under S1 and S2 in Simulation 2 are displayed in Table~\ref{size4.tab} and Table~\ref{size5.tab}, respectively, with the last row displaying their ARE values. We can get similar conclusions from these two tables. Firstly, it is apparent that all the tests proposed by \cite{gorecki2017multivariate} perform well in homoscedastic scenarios but tend to be considerably oversized in heteroscedastic cases. This behavior is expected, as these tests were specifically designed under the assumption of homoscedasticity. Secondly, while $T_{\ZZC}^{\B}$ outperforms the tests proposed by \cite{gorecki2017multivariate} in heteroscedastic cases, it also exhibits a slightly liberal behavior.  However, it is worth noting that $T_{\ZZC}^{\B}$ demonstrates better performance than those  presented in Table~\ref{size1.tab}. This is expected since this simulation study incorporates a larger number of observations and $T_{\ZZC}^{\B}$ can have a better performance if the sample size is large. Thirdly,
 $T_{\ZZC}^{\N}$ exhibits the best performance in homoscedastic scenarios and ranks second in heteroscedastic cases, whereas $T_{\NEW}$ demonstrates the opposite pattern, performing best in heteroscedastic scenarios and second in homoscedastic cases. Both of them perform reasonably well as their ARE values remain below 20, consistent with the benchmark established by \cite{zhang2012approximate}. Hence, it can be concluded that $T_{\NEW}$ remains effective even in scenarios where simulated functional samples exhibit measurement error or the simulated functional samples are sparse. 

\begin{table}[!h]  
	\caption{\em Simulation 2. Empirical sizes (in $\%$) of  W, LH, P, R, $T_{\ZZC}^{\N}$, $T_{\ZZC}^{\B}$, and $T_{\NEW}$ under S2. }  \label{size5.tab}	
		\begin{tabular*}{\textwidth}{@{\extracolsep{\fill}}cccccccccc}
			\toprule
            \multicolumn{9}{l}{Homoscedastic cases}\\
            \midrule
			$a$	& $\bn$        & W     & LH    & P     & R     & $T_{\ZZC}^{\N}$ & $T_{\ZZC}^{\B}$  & $T_{\NEW}$ \\
			\cmidrule{3-9}
    \multirow{3}[0]{*}{0.1} & $\bn_1$    & 5.6   & 5.6   & 5.6   & 5.5   & 4.4   & 8.0   & 3.7 \\
          & $\bn_2$    & 5.2   & 5.2   & 5.3   & 5.5   & 4.9   & 7.3   & 5.0 \\
          & $\bn_3$    & 5.5   & 5.6   & 5.4   & 5.2   & 4.5   & 7.3   & 4.6 \\
    \multirow{3}[0]{*}{0.5} & $\bn_1$    & 4.3   & 4.3   & 4.2   & 4.4   & 5.0   & 7.9   & 4.7 \\
          & $\bn_2$    & 5.3   & 5.4   & 5.4   & 5.1   & 5.2   & 7.2   & 5.5 \\
          & $\bn_3$    & 5.4   & 5.5   & 5.4   & 4.7   & 5.6   & 7.6   & 5.4 \\
    \multirow{3}[0]{*}{0.9} & $\bn_1$    & 6.1   & 6.0   & 5.8   & 5.9   & 5.2   & 8.9   & 5.3 \\
          & $\bn_2$    & 5.5   & 5.5   & 5.5   & 5.8   & 5.6   & 6.9   & 5.6 \\
          & $\bn_3$    & 4.8   & 4.9   & 4.8   & 4.6   & 5.4   & 6.7   & 5.2 \\
           \cmidrule{3-9}
    \multicolumn{2}{c}{ARE} & 10.0  & 10.2  & 9.8   & 9.6   & 7.1   & 50.7  & 8.9 \\

 \midrule
\multicolumn{9}{l}{Heterscedastic cases}\\
\midrule
  $a$	& $\bn$        & W     & LH    & P     & R     & $T_{\ZZC}^{\N}$ & $T_{\ZZC}^{\B}$  & $T_{\NEW}$ \\
			\cmidrule{3-9}
    \multirow{3}[0]{*}{0.1} & $\bn_1$    & 8.5   & 8.8   & 8.6   & 8.3   & 4.9   & 10.1  & 4.3 \\
          & $\bn_2$    & 8.6   & 8.4   & 8.6   & 8.3   & 5.8   & 8.0   & 5.3 \\
          & $\bn_3$    & 8.6   & 8.6   & 8.4   & 8.9   & 5.8   & 8.0   & 6.1 \\
    \multirow{3}[0]{*}{0.5} & $\bn_1$    & 10.7  & 11.1  & 10.4  & 11.2  & 6.5   & 10.0  & 5.4 \\
          & $\bn_2$    & 11.7  & 11.6  & 11.8  & 11.8  & 6.5   & 7.9   & 6.1 \\
          & $\bn_3$    & 11.2  & 11.1  & 10.9  & 10.4  & 5.7   & 7.4   & 5.4 \\
    \multirow{3}[0]{*}{0.9} & $\bn_1$    & 11.5  & 11.6  & 11.3  & 12.6  & 6.5   & 10.4  & 4.7 \\
          & $\bn_2$    & 10.6  & 10.7  & 10.6  & 10.4  & 5.7   & 7.6   & 5.0 \\
          & $\bn_3$    & 10.3  & 10.5  & 10.2  & 10.7  & 4.9   & 5.9   & 4.5 \\
           \cmidrule{3-9}
    \multicolumn{2}{c}{ARE} & 103.8 & 105.3 & 101.8 & 105.8 & 17.1  & 67.3  & 10.7 \\
\bottomrule
    \end{tabular*}%
\end{table}%

\section{Real Data Applications}\label{real.sec}
\begin{figure}[!h]
	\centering\includegraphics[width=\textwidth]{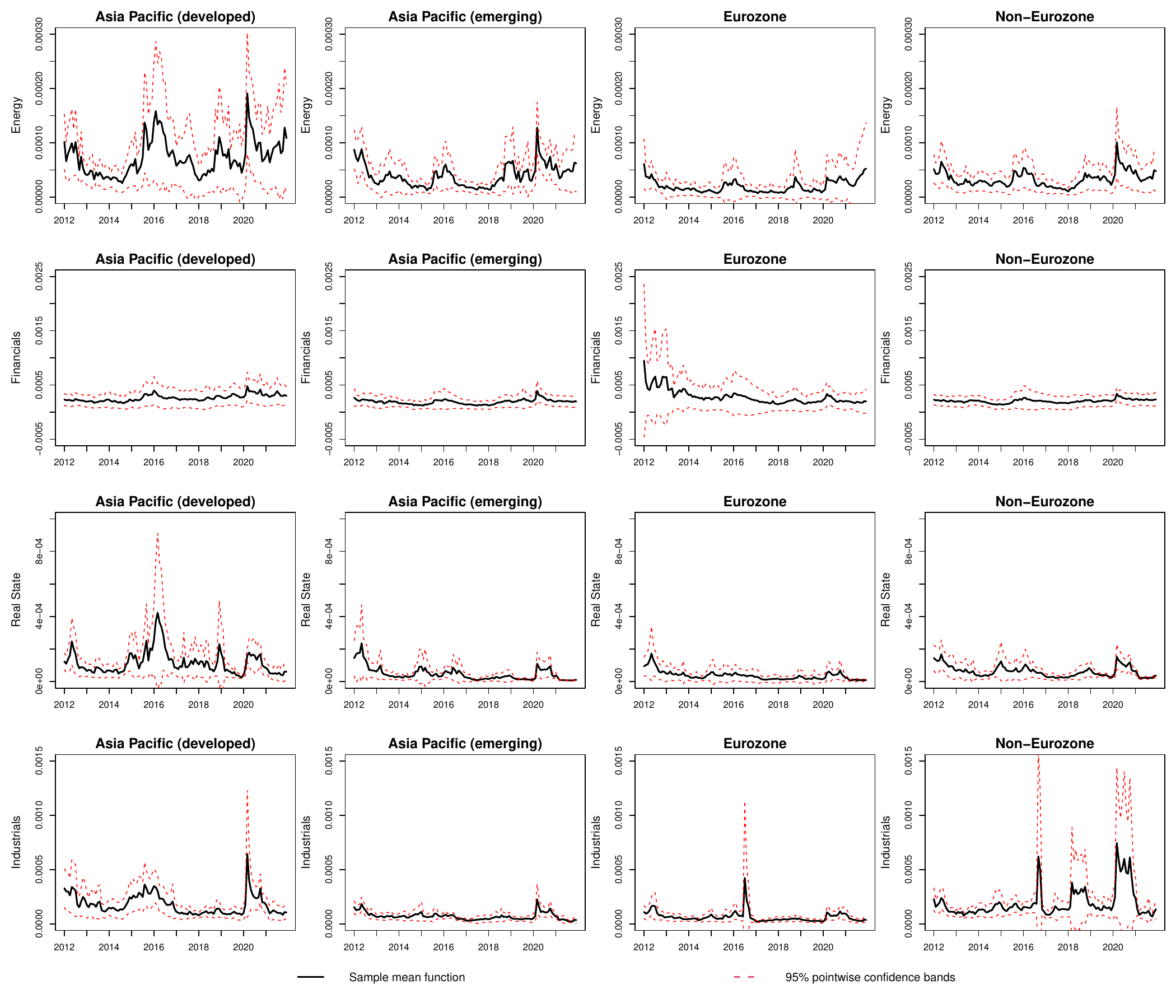}
	\caption{Group sample mean functions (solid) and their 95\% pointwise confidence bands (dashed) of the smoothed PD values of energy, financial, real estate, and industrial  of the firms	in R1: Asia Pacific (Developed),  R2: Asia Pacific (Emerging), R3: Eurozone, and R4: Non-Eurozone.}\label{real.fig}
\end{figure}

In this section, we apply  W, LH, P, R, $T_{\ZZC}^{\N}$, $T_{\ZZC}^{\B}$, and $T_{\NEW}$ to the financial data set which has been briefly described in Section~\ref{sec:intro}. This financial data set contains the  mean PD values aggregated by the economy of domicile and sector of each firm from 2012 to 2021. In particular, the number of firms in the four regions, that is, R1: Asia Pacific (Developed),  R2: Asia Pacific (Emerging), R3: Eurozone, and R4: Non-Eurozone are 7, 10, 11, and 14, respectively. It is of interest to compare the mean aggregated PD curves corresponding to four important factors, namely, energy, financial, real estate, and industrial in the four regions are all the same. In data preparation, since the observed discrete multivariate functional observations are measured in different time points, we reconstruct individual functions by smoothing splines and re-evaluate them at the same design time points. Fig.~\ref{real.fig} displays the pointwise sample group mean functions and their 95\% pointwise confidence bands of the four regions. 

To check whether the underlying mean functions of the smoothed PD values of energy, financial, real estate, and industrial  of the four regions are all the same,  we apply  W, LH, P, R, $T_{\ZZC}^{\N}$, $T_{\ZZC}^{\B}$, and $T_{\NEW}$. The test statistics and their corresponding $p$-values are shown in the first two rows of Table~\ref{pd.tab1}. Since all the considered tests suggest a strong rejection of the null hypothesis, we conclude that the underlying mean functions of the PD values of energy, financial, real estate, and industrial are unlikely the same for the four regions.
\begin{table}[!h]
		\caption{Testing results for one-way FMANOVA for the financial data set.}	\label{pd.tab1}
		\centering
		\begin{tabular*}{\textwidth}{@{\extracolsep{\fill}}lccccccc}
		\toprule
 & W     & LH    & P     & R 	& $T_{\ZZC}^{\N}$ & $T_{\ZZC}^{\B}$  & $T_{\NEW}$\\
	\midrule
	Test Statistic  &0.570 &0.647&0.494 &0.448 &$30.55$ &$30.55$ &$11.39$ \\
	$p$-value  &0.001 &0.001 &0.002 &0.001 &0.004 &$<0.001$ &$0.016$\\
    Empirical size (in \%) &9.2  &9.5  &9.1 &14.7 &11.5 &17.8  &5.8\\
   Empirical power (in \%) &87.8 &88.9 &85.7 &95.3 &98.5 &99.3 &96.3\\
	\bottomrule
	\end{tabular*}
\end{table}%

To verify the accuracy of the above testing results for the financial data set, we conducted a simulation study inspired by the approach of \cite{munko2024multiple} to access empirical sizes and powers.  The empirical sizes and powers are calculated as the proportion of times that the $p$-values are smaller than the nominal level $\alpha = 5\%$ based on the 1,000 simulation runs, and are displayed in Table~\ref{pd.tab1}. The setup for this study is as follows. In each iteration, we generated four multivariate functional samples, each with sample sizes corresponding to those of the firms in the four regions,  i.e., $n_1 = 7, n_2 = 10, n_3 = 11$, and $n_4 = 14$, drawn from Gaussian processes with group-wise covariance functions equal to the corresponding sample covariance functions. To calculate the empirical size, the mean function vector in each group is set to the vector of sample mean functions from the pooled data. For power investigation,  the mean function vector in the $i$-th group is set to the sample mean function vector for the $i$-th sample from the dataset.   As shown in Table~\ref{pd.tab1},  our proposed new test, $T_{\NEW}$, demonstrates better accuracy in maintaining the desired level ($\alpha=5\%$) compared to its competitors, and its power performance is at least on par with, if not superior to, the other tests.

\begin{table}[!h]
	\caption{$P$-values for some contrast tests  for the financial data.}\label{pd.tab2}
	\begin{tabular*}{\textwidth}{@{\extracolsep{\fill}}llllllll}
\toprule
		Hypothesis  & W     & LH    & P     & R 		& $T_{\ZZC}^{\N}$ & $T_{\ZZC}^{\B}$  & $T_{\NEW}$\\
	\midrule
		R1 vs. R2 &0.008 &0.004 &0.011 &0.002 
				&	$<0.001$	&	$<0.001$ &	$<0.001$\\
		R1 vs. R3 &0.705 &0.710 &0.704 &0.736 &0.367 &0.318 &0.554\\
		R1 vs. R4 &0.292 &0.266 &0.317 &0.041 
		&	$0.001$	&	$<0.001$ &	0.001\\
		R2 vs. R3 &$<0.001$ &$<0.001$ &0.001 &$<0.001$ 
		&$<0.001$	&	$<0.001$	&0.002\\
		R2 vs. R4 &0.007 &0.007 &0.007 &0.013 &0.063 &0.002 &0.135\\
		R3 vs. R4 &0.357 &0.350 &0.362 &0.223 &0.058 &$0.002$ &0.045\\
		\bottomrule
	\end{tabular*}
\end{table}%
Since the heteroscedastic one-way FMANOVA is highly significant, we further apply the tests under consideration to some contrast tests to check whether any two of the four regions have the same underlying groups mean functions of the PD values of energy, financial, real estate, and industrial. 
The results of these contrast tests are presented in Table~\ref{pd.tab2}, demonstrating largely consistent conclusions across all tests. Specifically, all tests agree that the contrast tests ``R1 vs. R2" and ``R2 vs. R3" yield significant results, while ``R1 vs. R3" does not, indicating no significant difference between the group mean functions of R1 and R3. However, divergent conclusions were observed for the contrast tests ``R1 vs. R4", ``R2 vs. R4", and ``R3 vs. R4". Let us examine the results of comparing R4 with either R1 or R3. In the case of ``R1 vs. R4", W, LH, and P failed to reject the null hypothesis, whereas R, $T_{\ZZC}^{\N}$, $T_{\ZZC}^{\B}$, and $T_{\NEW}$ rejected it at a significance level of 5\%. Similarly, for ``R3 vs. R4", W, LH, P, R, and $T_{\ZZC}^{\N}$ did not reject the null hypothesis at a significance level of 5\%, while $T_{\ZZC}^{\B}$ and $T_{\NEW}$ did. Notably, R and $T_{\ZZC}^{\N}$ yielded inconsistent results across the contrast tests ``R1 vs. R3", ``R1 vs. R4", and ``R3 vs. R4", suggesting potential reliability issues with these tests. Furthermore, as illustrated in Fig.~\ref{real.fig}, notable differences in mean functions are observed between regions R1 and R4, as well as between regions R3 and R4.  It is also important to note that the tests W, LH, P, and R were developed under the assumption of homoscedasticity, which may not hold true in this context where the covariance function matrices of different regions are likely to be distinct.  By contrast, our proposed new global test, $T_{\NEW}$, demonstrates superior size control compared to its competitors in the aforementioned simulation studies. Consequently, the $p$-values generated by $T_{\NEW}$ can be considered more reliable and trustworthy. 


\section{Concluding Remarks}\label{sec:conc}
In this paper, we propose and study a new global test for the general linear hypothesis testing problem for multivariate functional data. By adopting an adjustment coefficient,  our test controls the nominal size generally well and does not require a relative large sample size compared to \cite{zhu2023ksampFMD}'s tests. The null limiting distribution of the proposed test is approximated using the three-cumulant matched chi-squared-approximation. Some simulation studies demonstrate that the proposed test generally performs better or no worse than the existing tests for multivariate functional data in terms of size control, and performs well not only for dense functional data but also for sparse ones. Note that in hypothesis testing, maintaining accurate size control is the primary requirement before seeking better power performance (\citealt{li2023finite}). Hence,  the proposed test made an attempt in this direction. Unfortunately, Tables~\ref{size1.tab} and~\ref{size2.tab} also indicate that the proposed global test is still somewhat oversized when the
sample sizes are too small and the functional data are highly correlated. Methods for further improving the size control of the proposed test are interesting and warranted.  In addition, Table~\ref{size2.tab} demonstrates that the proposed test is less powerful than $T_{\QCZ}^{\max}$ when the functional data are highly or moderately correlated. Expending the current approach for power improvement provides another interesting direction for future work.

\mrk{In our analysis, Type I error is controlled for each contrast separately, but not across all hypotheses simultaneously. As one of the anonymous reviewers correctly pointed out, performing multiple tests increases the likelihood of falsely rejecting at least one null hypothesis. While our current approach does not apply formal multiple testing corrections, methods such as the Bonferroni correction, as used by \cite{munko2024multiple} for \cite{zhu2023ksampFMD}'s test, could be considered to control the family-wise error rate (FWER). However, based on their simulation studies, simpler corrections may not always be the most effective choice. Future work could explore incorporating more sophisticated corrections into our framework to ensure robust inference across multiple hypotheses.}

\backmatter

 \bmhead{Supplementary information}
The R code used in this study is provided in the Supplementary Material. 



 \bmhead{Acknowledgements}
 This research was supported by the National Institute of Education (NIE), Singapore, under its start-up grant (NIE-SUG 6-22 ZTM). The computational work for this article was partially performed on resources of the National Supercomputing Centre, Singapore (\url{https://www.nscc.sg}) with project IDs 12002633 and 12003037. The author thanks two anonymous reviewers for their constructive comments and suggestions, which helped improve the article substantially, and Dr. Zhiping Qiu for providing the code from \cite{Qiu2021}.









\begin{appendices}

\section{Technical Proofs}\label{proof.sec}

\setcounter{equation}{0}
\global\long\def\theequation{A.\arabic{equation}}
\begin{proof}[Proof of Theorem~\ref{T0.thm}.]

	Let $\barbx_i(t),i=1,\ldots,k$ be the usual vector of sample mean functions of $\bx_{ij}, j=1,\ldots,n_i;\;i=1,\ldots,k$ which are defined in (\ref{xij.equ}).  From (\ref{Tnde.sec2}), we can rewrite $T_{n,0}^*$ as 
    \begin{equation}\label{Tn0.equA}
    \begin{split}
          T_{n,0}^* &= \int_{\calT}\tr\left[\bOmega_n^{-1/2}(t,t)\bX(t)^\top\bbH\bX(t)\bOmega_n^{-1/2}(t,t)\right]dt\\
          &=\sum_{i=1}^k\sum_{j=1}^kh_{ij}<\barbx_i^*,\barbx_j^*>_{\HS}=\int_{\calT}\barbx^{*}(t)^\top(\bbH\otimes\bI_p)\barbx^{*}(t)dt,
    \end{split}
    \end{equation}
    where  $\barbx^*(t)=[\barbx^*_1(t)^\top,\ldots,\barbx^*_k(t)^\top]^\top$,  a long column vector of dimension $kp$, obtained by stacking all sample mean vectors  $\barbx^*_1(t),\ldots,\barbx^*_k(t)$ with $\barbx^*_i(t)=\bOmega_n^{-1/2}(t,t)\barbx_i(t),i=1,\ldots,k$. Setting $\bC=[(\bG\bD\bG^\top)^{-1/2}\bG]\otimes \bI_p$, we have $\bC^\top\bC=\bbH\otimes \bI_p$, allowing us to further rewrite $T_{n,0}^*$  in (\ref{Tn0.equA}) as $T_{n,0}^*=\|\bC\barbx^*\|_{\HS}^2$.
    
  As $n_{\min}\to \infty$, by the central limit theorem of i.i.d. stochastic processes (\citealt{van1996weak}), we have $\barbx_i(t)\convL \GP_{p}(\bzero, \bGamma_i/n_i)$, where $\GP_p(\bm{\eta},\bGamma)$ denotes a $p$-dimensional Gaussian process with vector of mean functions $\bm{\eta}(t)$ and matrix of covariance functions $\bGamma(s,t)$, and similarly we have $\barbx^*_i(t)\convL \GP_{p}(\bzero, \bGamma^*_i)$, where $\bGamma^*_i(s,t)=\bOmega_n^{-1/2}(s,s)\bGamma_i(s,t)\bOmega_n^{-1/2}(t,t)/n_i,i=1,\ldots,k$. Since the $k$ samples are independent, it follows that $\bC\barbx^*(t)\convL \GP_{qp}(\bzero,\bSigma)$ where $\bSigma(s,t)=\bC\diag[\bGamma_1^*(s,t),\ldots,\bGamma_k^*(s,t)]\bC^\top$.
 
 By applying Lemma 3 in \cite{zhu2023ksampFMD},  under Conditions {C1--C3},  $T_{n,0}^*$, which is the squared $L^2$-norm of $\bC\barbx^*(t)$ can be expressed as $	T_0^*\convd \sum_{r=1}^\infty \lambda_rA_r$,
	where	$A_1,A_2,\ldots \iidsim \chi_1^2$ and $\lambda_1,\lambda_2,\ldots$ are  eigenvalues of $\bSigma(s,t)$ in descending order. Let $\bm{\phi}_{r}(t),r=1,2,\ldots,$ denote the eigenfunctions of $\bSigma(s,t)$ corresponding to $\lambda_r,r=1,2,\ldots.$ It follows that $\calK_1(T_0^*)=\sum_{r=1}^\infty\lambda_r=\tr(\bSigma)$, $\calK_2(T_0^*)=2\sum_{r=1}^\infty\lambda_r^2=2\tr(\bSigma^{2})$, and $\calK_3(T_0^*)=8\sum_{r=1}^\infty \lambda_r^3=8\tr(\bSigma^{3})$. The last equality is given by
	\begin{equation*}
		\begin{array}{rcl}
		&&\quad	\tr(\bSigma^{3})=\int_{\calT^3}\tr[\bSigma(s,t)\bSigma(t,v)\bSigma(v,s)]dsdtdv\\
			&=&	\int_{\calT^3}\tr\Big\{\Big[\sum\limits_{r_1=1}^\infty\lambda_{r_1} \bm{\phi}_{r_1}(s)\bm{\phi}_{r_1}(t)^{\top}\Big]\Big[\sum\limits_{r_2=1}^\infty\lambda_{r_2} \bm{\phi}_{r_2}(t)\bm{\phi}_{r_2}(v)^{\top}\Big]\Big[\sum\limits_{r_3=1}^\infty\lambda_{r_3} \bm{\phi}_{r_3}(v)\bm{\phi}_{r_3}(s)^{\top}\Big]\Big\}dsdtdv\\
			&=&\sum\limits_{r_1=1}^\infty\sum\limits_{r_2=1}^\infty\sum\limits_{r_3=1}^\infty\lambda_{r_1}\lambda_{r_2}\lambda_{r_3}\tr\Big[\int_{\calT^3}\bphi_{r_1}(s)\bphi_{r_1}(t)^\top\bphi_{r_2}(t)\bphi_{r_2}(v)^\top\bphi_{r_3}(v)\bphi_{r_3}(s)^\top dsdtdv\Big]\\
			&=&\sum_{r=1}^\infty \lambda_r^3.
		\end{array}
	\end{equation*}
	By some simple algebra, we have 
    \[
\begin{array}{rcl}
	\calK_1(T_0^*)
	&=&\tr(\bSigma) = \int_{\calT}\tr\left\{(\bbH\otimes \bI_p) \diag[\bGamma_1^*(t,t),\ldots,\bGamma_k^*(t,t)]\right\}dt\\
    &=&\int_{\calT}\sum_{i=1}^kh_{ii}\tr[\bGamma_i^*(t,t)]dt=\int_{\calT}\sum_{i=1}^kh_{ii}\tr[\bOmega_n^{-1}(t,t)\bGamma_i(t,t)/n_i]dt\\
    &=&\int_{\calT}\tr[\bOmega_n^{-1}(t,t)\sum_{i=1}^k h_{ii}\bGamma_i(t,t)/n_i]dt=\int_{\calT}\tr(\bI_p)=(b-a)p.
    \end{array}
    \]
   For simplicity of notation, let $\Upsilon(s,t)=(\bbH\otimes \bI_p) \diag[\bGamma_1^*(s,t),\ldots,\bGamma_k^*(s,t)]$, then we have
      \[
\begin{array}{rcl}
	\calK_2(T_0^*)&=&2\tr(\bSigma^{2})=\int_{\calT^2}\tr[\Upsilon(s,t)\Upsilon(t,s)]dsdt\\
    &=&2\sum_{i=1}^k\sum_{j=1}^kh_{ij}h_{ji}\int_{\calT^2}\tr[\bGamma_i^*(s,t)\bGamma_j^*(t,s)]dsdt\\
    &=&2\sum_{i=1}^k\sum_{j=1}^kh_{ij}^2\tr(\bGamma_i^*\bGamma_{j}^*),\mbox{ and }\\
	\calK_3(T_0^*)
	&=&8\tr(\bSigma^{3})=\int_{\calT^3}\tr[\Upsilon(s,t)\Upsilon(t,v)\Upsilon(v,s)]dsdtdv\\
    &=&8\sum_{i=1}^k\sum_{j=1}^k\sum_{\ell=1}^kh_{ij}h_{j\ell}h_{\ell i}\int_{\calT^3}\tr[\bGamma_i^*(s,t)\bGamma_{j}^*(t,v)\bGamma_{\ell}^*(v,s)]dsdtdv\\
    &=&8\sum_{i=1}^k\sum_{j=1}^k\sum_{\ell=1}^kh_{ij}h_{j\ell}h_{\ell i}\tr(\bGamma_i^*\bGamma_{j}^* \bGamma_{\ell}^*).
\end{array}
\]
	The proof is then completed. 
\end{proof}

\begin{proof}[Proof of Theorem~\ref{T1.thm}.]
	When $n_{\min}$ is large, we have $T_n/c_n=T_n^*[1+o(1)]$. Together with the local alternative (\ref{H1.sec3}) and the decomposition of $T_{n}^*$ in (\ref{Tnde.sec2}), this yields
$T_n/c_n=\Big\{T_{n,0}^*+\int_{\calT}\tr\Big[\bM(t)^\top\bbH\bM(t)\bOmega_n^{-1}(t,t)\Big]dt\Big\}[1+o(1)].$
Theorem~\ref{T0.thm} indicates that as $n_{\min}\to\infty$, $T_{n,0}^*\convL \beta_0+\beta_1\chi_d^2$, then we have 
\[
\begin{split}
	&\quad\Pr\left[T_{n}/c_n\geq \hbeta_0+\hbeta_1\chi_{\hd}^2(\alpha)\right]\\
	&=\Pr\Big\{T_{n,0}^*\geq \hbeta_0+\hbeta_1\chi_{\hd}^2(\alpha)-\int_{\calT}\tr\Big[\bM(t)^\top\bbH\bM(t)\bOmega_n^{-1}(t,t)\Big]dt\Big\}[1+o(1)]\\
	&=\Pr\Big\{\frac{\chi_{d}^2-d}{\sqrt{2d}}\geq \frac{\chi_{d}^2(\alpha)-d}{\sqrt{2d}}-\frac{n\int_{\calT}\tr\left[\bM(t)^\top\bbH^*\bM(t)\bOmega^{-1}(t,t)\right]dt}{\sqrt{2	\sum_{i=1}^k\sum_{j=1}^kh_{ij}^{*2}\tr(\tilde{\bGamma}^*_{i}\tilde{\bGamma}^*_{j})}}\Big\}[1+o(1)],\\
\end{split}
\]
where $\bbH^*$, $\bOmega(t,t),t\in\calT$, and $\tilde{\bGamma}^*_{i},i=1,\ldots,k$ are defined in (\ref{limitOmegan.sec2}).

\end{proof}

\end{appendices}


 \bibliography{TSHTMFDref}

\end{document}